\documentclass{nature}

\usepackage{longtable}
\usepackage{graphicx}

\usepackage{times}
\usepackage{multirow}

\usepackage{amsmath}
\usepackage{wasysym}
\usepackage{graphicx}
\usepackage{rotating}
\usepackage[font=singlespacing]{caption}

\title{A primordial origin for the composition similarity between the Earth and the Moon}

\author
{Alessandra Mastrobuono-Battisti$^{1}$, Hagai B. Perets$^{1}$ \& Sean N. Raymond$^{2,3}$\\
}

\begin{document} 




\maketitle 

\begin{affiliations}
 \item Department of Physics, Technion, Israel Institute of Technology, Haifa, 32000
 \item CNRS, Laboratoire d'Astrophysique de Bordeaux, UMR 5804, F-33270, Floirac, France\\
  $^{3}$Univ. Bordeaux, Laboratoire d'Astrophysique de Bordeaux, UMR 5804, F-33270, Floirac, France
\end{affiliations}

\begin{abstract}

Most of the properties of the Earth-Moon system can be explained by a collision between a planetary embryo and the growing Earth late in the accretion process\cite{Ca01,ACL99,JM14}. Simulations show that most of the material that eventually aggregates to form the Moon originates from the impactor\cite{Ca01,Ca04,Ca08}.   However, analysis of the terrestrial and lunar isotopic composition show them to be highly similar\cite{Ri86,Lu98, Wi01, TK07, ZD12,HP14}. In contrast, the compositions of other solar system bodies are significantly different than the Earth and Moon\cite{CM96,Fr99,As14}. This poses a major challenge to the giant impact scenario since the Moon-forming impactor is then thought to also have differed in composition from the proto-Earth.  Here we track the feeding zones of growing planets in a suite of simulations of planetary accretion\cite{Ra09}, in order to measure the composition of Moon-forming impactors.  We find that different planets formed in the same simulation have distinct compositions, but the compositions of giant impactors are systematically more similar to the planets they impact.  A significant fraction of planet-impactor pairs have virtually identical compositions. Thus, the similarity in composition between the Earth and Moon could be a natural consequence of a late giant impact.  
\end{abstract}

Successful models for Moon formation typically require a relatively low-velocity, oblique impact \cite{Ca01} between the proto-Earth and up to a few$\times0.1$ Earth-mass ($M_E$) planetary embryo. Such Moon-forming impacts typically occur at the late stages of planetesimal accretion by the terrestrial planets\cite{ACL99,JM14}. A circum-terrestrial debris disk is formed from material ejected during these impacts. 
The composition of the disk, in a typical impact, is dominated by material from the impactor mantle ($>60$ weight percent, \% wt\cite{Ca01,Ca04,Ca08}) with a smaller contribution (typically $\approx20\%$) from the proto-Earth. 
 More material can be extracted from the proto-Earth when a slightly sub-Mars sized body hits a fast spinning planet, that is later slowed down by resonances\cite{Cu12}. The spin should be close to the break-up velocity. Another possible channel producing a significant mixed material from both the planet and impactor is the rare collision between two comparable mass embryos,  which both masses are of about half of Earth's mass\cite{Ca12}. Although these new models can potentially solve some of the composition issues borne by the giant-impact scenario, they do require ad hoc assumptions and pose several difficulties (see ref. \cite{JM14} for a discussion). Here, we focus on the former typical giant-impact events,  in which the Moon aggregates mostly from material originating from the impactor. 
 
Lunar meteorites and rock samples returned by the Apollo mission have a very similar composition to Earth's mantle, across a variety of different isotopes\cite{Ri86,Lu98, Wi01, TK07, ZD12,HP14}. Combining these with the giant-impact simulation results, one infers that the Moon-forming impactor and the Earth should have had a similar composition. This poses a fundamental difficulty to the giant-impact model for the origin of the Moon, since analysis of material from other solar system bodies have shown them to significantly differ from that of the Earth (see refs. \cite{CM96,Fr99} and ref. \cite{As14} for a review). This would suggest that the composition of the Moon-forming impactor should have similarly differed from that of the Earth, in contrast with the giant-impact basic prediction. 

Here, we analyze the results of extensive N-body simulations of terrestrial planet formation to show that the Earth-Moon composition challenge can be alleviated. In particular, we show that the compositions of a significant fraction, 20\% to 40\%, of giant-impactors are consistent with being similar to that of the planets they impact.  More generally, late giant impactors have significantly more similar compositions to the planets they impact compared with other planets in the same system, showing large differences. 
 
To study the compositions of planets and their impactors we analyzed 40 dynamical simulations (from ref. \cite{Ra09}; using the Mercury code\cite{Ch99}) of the late stages of planetary accretion, following the formation of Jupiter and Saturn, and after all the gas in the protoplanetary disk has been dissipated and/or accreted to the gas-giants (see ref. \cite{Ra13} for a recent review).  Each simulation started from a disk of 85-90 planetary embryos and 1000-2000 planetesimals  extending from $0.5$ to $4.5$~AU. Jupiter and Saturn are fully formed and have different orbits and inclinations in different sets of simulations  (detailed descriptions can be found in  ref. \cite{Ra09} and in the Methods. 
Within 100-200 million years, each simulation typically produced 3-4 rocky planets formed from collisions between embryos and planetesimals. Each of these planets accreted a large number of planetesimals during its evolution. All collisions were recorded and provide a map of each planet's feeding zone (see also ref. \cite{Ra06}).  Assuming that the initial composition of material in the protoplanetary disk is a function of its position in the initial protoplanetary disk, one can compare the compositions of different bodies formed and evolved in the simulations.     

Previous studies explored the compositions of the different planets formed in similar simulations. However, the composition of impactors on formed planets have been hardly explored. Pahlevan \& Stevenson\cite{PS07} analyzed a single statistically limited simulation which included a total of $\sim$150 particles, and compared the compositions of any impactors on any planets during the simulation (not only giant impacts, due to small number statistics) to that of the planets. They concluded that the scatter among the compositions of the various impactors is comparable to the observed differences between the planets. In particular, they found that none of the planetary impactors in the simulation they analyzed had an isotopic composition similar enough to the final planet to yield an Earth-Moon-like composition similarity. 

Using the data from our large set of high resolution simulations, we compare the composition of each surviving planet with that of its last giant impactor, i.e. the last planetary embryo that impacted the  planet (typical impactor-to-planet mass ratio in the range $\approx0.2-0.5$; see Table 1).  We only include  the 20 cases where both the impactor and planet are composed of at least 50 particles each, so as to have significant statistics. Analysis of the additional data for impactors composed from a smaller number of particles (and hence smaller statistics for the specific composition) are consistent with the higher resolution cases discussed here, as shown in the Methods.  The comparison is then done as follows. First we compare the feeding zones of the planet and impactor, as shown in the examples in Fig. 1 (the cumulative plots for these and all other cases can be found in the Methods). We calculate the probability (P) that the feeding zones of the impactor and the planet are drawn from the same distribution, using a two-group Kolmogorov-Smirnov (KS) test (probabilities shown in the plots and in Table 1).  In 3 out 20 cases the feeding zones contributing to the Moon and those contributing to the planet are consistent with being drawn from the same parent distribution.  In other words, the Moon feeding zones, if derived solely from the impactor, are consistent with the Earth's in $15\%$ of the impacts. The consistency further improves if we assume that a fraction of the proto-Earth was mixed into the  Moon (as suggested by detailed collision simulations showing a $10-40\%$ contribution from the proto-Earth\cite{As14}).  For the typical $20\%$ mix of proto-Earth material with the impactor material forming the Moon (as found in simulations), $35\%$ of  cases are consistent, and the success rate increases further for a higher mass contribution from the proto-Earth (see Table 1 and Extended Data Figures 1 and 2). While this shows that the proto-Earth and the Moon-forming impactor may have had similar feeding zones, it does not yet quantitatively guarantee that the composition is as similar as that of the Earth-Moon system.

We therefore further explore the compositional similarities with the Earth-Moon system, and calculate the oxygen isotope ratios of our simulated planets.  We assume that a linear gradient existed in the $^{17}{\rm O}$ isotopic composition in the initial protoplanetary disk of the solar-system.  Following Pahlevan \& Stevenson\cite{PS07} we calibrate the initial $^{17}{\rm O}$ isotopic composition in each of our simulations using Earth and Mars' measured compositions (see ref. \cite{PS07} and the Methods for more details;  where we also discuss the sensitivity of results to the calibration used, as well as the criteria for which planet-impactor pairs are considered in the analysis. We find qualitatively similar results when using different criteria and calibrations, as we discuss in detail in the Methods).  Given this calibration we assign each planetesimal and planetary embryo a specific initial $^{17}{\rm O}$ isotopic abundance based on its initial orbit, and then average the contribution of all accreted planetesimals, while weighting each accreted planetesimal/embryo according to its mass,  to obtain the $^{17}{\rm O}$  of the planets and impactors  and derive the offset between them.

The  $^{17}{\rm O}$ isotope is chosen for comparison since it provides the most stringent constraint on the Earth-Moon similarity, and its abundances were measured across a variety of solar system bodies (enabling the best opportunity for calibration). The measured difference between Mars' and Earth's $^{17}{\rm O}$ abundances (used for calibration) is $\Delta^{17}{\rm O}_{\rm Mars}=+321\pm13$ parts per million (ppm)\cite{HP14} (a similarly large difference was found for the composition of 4 Vesta asteroid derived from HED meteorites; $-250\pm60$ ppm).  The difference between the Earth and Moon is just $\Delta^{17}{\rm O}=12\pm3$ ppm\cite{HP14}.  In Table 1 we show the $\Delta^{17}{\rm O}$ differences between the impactors and the planets in our simulations.  Note that we adopt the same Earth-Mars composition difference calibration as used by Pahlevan \& Stevenson\cite{PS07}.  The results are linearly dependent on the adapted calibration; see Table 1. The calibration factor is defined by C$_{\rm cal}=(\Delta^{17}{\rm O}_4-\Delta^{17}{\rm O}_3)/\Delta^{17}{\rm O}_{\rm Mars}$, where $\Delta^{17}{\rm O}_4$ and  $\Delta^{17}{\rm O}_3$ refer to the compositions of the fourth and third planet in the simulations (unless only three planets formed, in which case the third and second planet were taken for calibration purposes).
 
In $20\%$ 
of the cases the impactors and planets have absolute offsets comparable or smaller than the measured absolute offset in the Earth-Moon system, i.e. smaller than the 1$\sigma$ limit estimated using the Lunar samples ($<15$ppm). Taking into account the 1$\sigma$ uncertainty calculated for the $\Delta^{17}O$ in the simulated systems, the fraction of consistent pairs can be as large as 40\% of the whole sample. This fraction becomes larger when partial mixing of Earth material is allowed, as observed in simulation data (it increases to $50\%$ ($55\%$), for $20\%$ ($40\%$)  contribution from the planet; see Table 1 and Fig. 2).  Even planet-impactor pairs with statistically different feeding zones have $\Delta^{17}{\rm O}$ offsets significantly smaller than those found for Mars and Vesta, in most cases. More generally, planet-impactor pairs are robustly more similar in composition compared with pairs of surviving planets in the same system (see Fig. 2, Extended Data Figures 3, 4 and 5, as well as the Supplementary Information Table, for the $\Delta^{17}{\rm O}$ difference distribution for planets and impactors).  Not less important, the differences between the planets are of similar order to those found between the Earth, Mars and Vesta, i.e. consistent with the observations of the solar system. Interestingly, a small fraction of the planets do have very similar composition (small $\Delta^{17}{\rm O}$ difference), suggesting the possibility for the existence of solar system bodies with similar compositions to the Earth besides the Moon. 
As shown in Extended Data Table 2 and in the Methods this result still holds when considering lower thresholds for the minimal number of particles composing the planet/impactor (between 1 to 40). In particular, the mean fraction of compatible planet-impactor pairs extends between $10$\% and $20$\% for all cases. The fractions become even higher ($20$\%-$40$\%) when accounting for the $1\sigma$ uncertainties.

The Earth-Moon composition similarity poses a major challenge to the standard model of the giant-impact scenario as it conflicts with the predominant derivation of the Moon composition from the mantle of the impacting planet\cite{As14}. It therefore gave rise to a wide range of alternative impact scenarios\cite{Be05,PS07,Ca12,Cu12,Sa12,Re12,As14}.
However, all of these models suffer some potentially considerable difficulties and/or require fine-tuned conditions (see ref. \cite{As14} for a review). Our analysis of solar-system-like planet formation scenarios  potentially offers a solution to the major composition-similarity obstacle, for the standard giant-impact scenario. We find that a significant fraction of all planetary impactors could have had similar composition to the planets they impacted, in contrast with the composition of different planets existing at the same planetary system.  Note that the solution of impactor-planet similar composition suggested by our results well applies for the origin of the  $\Delta^{17}{\rm O}$ similarity between the Earth and the Moon, and may similarly apply for the other isotopic similarities of the Silicon and Tungsten\cite{2014RSPTA.372.0244D}. However, it is still debated whether even a similar impactor-planet composition could resolve the composition similarity of the latter (e.g. Silicon), as Earth's silicate mantle may reflect the consequences of silicon sequestration by a core formed at high temperatures on a large planetary body\cite{el+13,2014RSPTA.372.0244D}, i.e. larger than the typical impactors considered.

  We conclude that our findings  can potentially resolve the apparent contrast between the observed similarity of the Earth and the Moon composition and its difference from that of other solar system bodies. This primordial composition similarity solution may therefore lift the prime obstruction for the standard giant-impact origin of the Moon, as well as ease some of the difficulties for alternative giant impact scenarios suggested in recent years. 

\bibliography{moon}


\begin{addendum}
 \item HBP acknowledge support from BSF grant number 2012384, the Minerva center for life under extreme planetary conditions and the ISF I-CORE grant 1829/12. We would like to thank Oded Aharonson for helpful remarks on early version of this manuscript. We thank Nathan Kaib for the useful discussion on 
 related work.
 \item[Supplementary Information] is available in the online version of the paper.
 \item[Author Contributions] AMB analyzed the simulation data and produced the main results; HBP initiated and supervised the project and took part in the data analysis. SNR provided the simulation data used for the analysis. The paper writing was lead by AMB and HBP with contributions from SNR.
 \item[Competing Interests] The authors declare that they have no
competing financial interests.
 \item[Correspondence] Correspondence and requests for materials
should be addressed to A.M.B. and H.B.P. ~(email: amastrobuono@physics.technion.ac.il, hperets@physics.technion.ac.il).
\end{addendum}

\clearpage
\setcounter{table}{0}
\begin{sidewaystable} \tiny
	\begin{tabular}{|c|c|c|c|c|c|c|c|c|c|c|c|c|c|}
		\hline 
		\textbf{Model } & \textbf{\#} & \textbf{M$_{{\rm P}}$ } & \textbf{M$_{{\rm I}}$ } & \textbf{$\frac{M_{{\rm I}}}{M_{{\rm P}}}$} & \textbf{N$_{{\rm P}}$ } & \textbf{N$_{{\rm I}}$ } & \textbf{$t_{{\rm coll}}$ } & \multicolumn{3}{c|}{\textbf{${\rm C_{{\rm cal}}}\Delta^{17}O$ (ppm)}} & \multicolumn{3}{c|}{\textbf{KS-probability}}\tabularnewline
		\cline{9-14} 
		&  & \textbf{($M_{\oplus}$) } & \textbf{($M_{\oplus}$) } &  &  & \textbf{Â  } & \textbf{(Myrs)} & \textbf{$0\,\%$} & \textbf{$20\,\%$} & \textbf{$40\,\%$} & \textbf{$0\,\%$} & \textbf{$20\,\%$} & \textbf{$40\,\%$}\tabularnewline
		\hline 
		cjs15  & 1 & $0.94$ & $0.43$ & $0.46$ & $123$ & $97$ & $50.7$ & $13\pm14$ &  $10\pm13$ &  $8\pm12$ & $0.0039$ & $0.13$ & $0.67$\tabularnewline
		\hline 
		cjs15  & 2 & $0.78$ & $0.27$ & $0.35$ & $209$ & $78$ & $80.9$ & $(-1.05\pm0.26)\times10^{2}$  & $-84\pm23$ & $-63\pm22$ & $1.3\times10^{-7}$ & $3.1\times10^{-4}$ & $0.079$\tabularnewline
		\hline 
		cjs1  & 3 & $1.25$ & $0.42$ & $0.34$ & $219$ & $73$ & $149.7$ & $64\pm13$ & $52\pm11$ &  $39.9\pm9.6$ & $5.3\times10^{-11}$ & $8.1\times10^{-7}$ & $0.0092$ \tabularnewline
		\hline 
		cjs1   & 4 & $1.05$ & $0.39$ & $0.37$ & $128$ & $78$ & $186.2$ & $(-1.97\pm0.20)\times10^{2}$  &  $(-1.57\pm0.20)\times10^{2}$ & $(-1.18\pm0.20)\times10^{2}$ & $1.1\times10^{-29}$ & $5.1\times10^{-18}$ & $0.052$ \tabularnewline
		\hline 
		cjs1   & 5$^{*}$  & $1.21$ & $0.38$ & $0.31$ & $219$ & $75$ & $123.5$ & $-24\pm17$  & $-20\pm15$ & $-15\pm13$ & $0.0023$ & $0.056$ & $0.19$\tabularnewline
		\hline 
		cjsecc  & 6 & $0.94$ & $0.36$ & $0.38$ & $117$ & $79$ & $80.4$ & $-51\pm34$  & $-40\pm30$  & $-31\pm27$  & $3.4\times10^{-4}$ & $0.025$ & $0.33$\tabularnewline
		\hline 
		cjsecc  & 7 & $1.01$ & $0.32$ & $0.32$ & $144$ & $68$ & $75.5$ & $-12\pm16$ &  $-10\pm13$ & $-7\pm12$ & $0.038$ & $0.050$ & $0.20$\tabularnewline
		\hline 
		cjsecc  & 8  & $1.02$ & $0.42$ & $0.41$ & $148$ & $89$ & $36.9$ & $13\pm79$ &  $11\pm71$ & $8\pm66$ & $0.054$ & $0.21$ & $0.53$\tabularnewline
		\hline 
		eejs15  & 9 & $0.70$ & $0.19$ & $0.27$ & $111$ & $52$ & $77.1$ & $9.1\pm7.4$  &  $7.3\pm6.2$ & $5.5\pm5.2$ & $0.071$ & $0.18$ & $0.32$\tabularnewline
		\hline 
		eejs15 & 10 & $0.55$ & $0.13$ & $0.24$ & $263$ & $65$ & $24.6$ & $98\pm29$  & $78\pm24$ & $59\pm22$ & $1.6\times10^{-12}$ & $2.3\times10^{-8}$ & $0.0025$\tabularnewline
		\hline 
		eejs15 & 11 & $0.78$ & $0.22$ & $0.29$ & $256$ & $69$ & $102.3$ & $(-1.08\pm0.35)\times10^{2}$ & $-87\pm31$ & $-65\pm27$ & $2.9\times10^{-13}$ & $7.9\times10^{-7}$ & $0.0069$\tabularnewline
		\hline 
		eejs15 & 12 & $0.73$ & $0.26$ & $0.36$ & $298$ & $87$ & $105.6$ & $26\pm18$  & $21\pm16$  & $16\pm14$  &$1.3\times10^{-10}$ & $5.5\times10^{-6}$ & $0.034$\tabularnewline
		\hline 
		eejs15 & 13$^{*}$ & $1.30$ & $0.33$ & $0.25$ & $525$ & $126$ & $199.8$ & $93\pm1.5\times10^{2}$ & $74\pm1.3\times10^{2}$ & $55\pm1.1\times10^{2}$  & $7.1\times10^{-8}$ & $5.4\times10^{-6}$ & $0.0043$\tabularnewline
		\hline 
		eejs15 & 14 & $0.50$ & $0.18$ & $0.36$ & $170$ & $53$ & $33.1$ & $-26\pm75$ & $-21\pm65$ & $-16\pm56$ & $0.14$ & $0.13$ & $0.24$\tabularnewline
		\hline 
		eejs15 & 15 & $0.50$ & $0.13$ & $0.26$ & $234$ & $61$ & $32.0$ & $-73\pm56$ & $-58\pm48$ & $-44\pm43$ & $2.7\times10^{-8}$ & $1.8\times10^{-7}$ & $0.062$\tabularnewline
		\hline 
		eejs15 & 16 & $0.67$ & $0.33$ & $0.49$ & $213$ & $120$ & $145.0$ & $(1.37\pm0.51)\times10^{2}$ & $(1.10\pm0.45)\times10^{2}$ & $83\pm42$ & $2.4\times10^{-5}$ & $0.013$ & $0.43$\tabularnewline
		\hline 
		eejs15  & 17$^{*}$  & $1.15$ & $0.41$ & $0.36$ & $177$ & $69$ & $168.3$ & $(7.9\pm1.3)\times10^{2}$ & $(6.3\pm1.1)\times10^{2}$ & $(4.75\pm0.93)\times10^{2}$ & $3.5\times10^{-19}$ & $1.2\times10^{-10}$ & $0.0028$\tabularnewline
		\hline 
		ejs15  & 18 & $0.81$ & $0.36$ & $0.44$ & $63$ & $55$ & $76.3$ & $-81\pm48$  & $-65\pm42$  & $-49\pm37$  & $1.5\times10^{-5}$ & $0.0039$ & $0.29$\tabularnewline
		\hline 
		jsres  & 19  & $1.04$ & $0.32$ & $0.31$ & $166$ & $89$ & $79.5$ & $55\pm22$ &  $44\pm19$ & $33\pm17$ & $3.0\times10^{-10}$ & $1.5\times10^{-6}$ & $0.0078$\tabularnewline
		\hline 
		jsres & 20  & $1.27$ & $0.63$ & $0.50$ & $134$ & $89$ & $176.8$ & $-84\pm29$ & $-67\pm25$ & $-50\pm23$ & $5.6\times10^{-7}$ & $0.011$ & $0.50$\tabularnewline
		\hline 
		\textbf{Obs.} &  &  & \multicolumn{5}{c|}{} & \multicolumn{3}{c|}{\textbf{$\Delta^{17}O$ (ppm)}} & \multicolumn{3}{c|}{}\tabularnewline
		\hline 
		Earth &  & $1$ & \multicolumn{5}{c|}{} & \multicolumn{3}{c|}{$0\pm3$} & \multicolumn{3}{c|}{}\tabularnewline
		\hline 
		Moon &  & $0.012$ & \multicolumn{5}{c|}{} & \multicolumn{3}{c|}{$12\pm3$} & \multicolumn{3}{c|}{}\tabularnewline
		\hline 79
		Mars &  & $0.07$ & \multicolumn{5}{c|}{} & \multicolumn{3}{c|}{$321\pm13$} & \multicolumn{3}{c|}{}\tabularnewline
		\hline 
		4 Vesta &  & $4.33\times10^{-5}$ & \multicolumn{5}{c|}{} & \multicolumn{3}{c|}{$-250\pm80$} & \multicolumn{3}{c|}{}\tabularnewline
		\hline 
	\end{tabular}
	\\$^{*}$ 3-planet systems; calibration was done on the 1$^{\rm st}$ and 2$^{\rm nd}$ planets.
	\caption{\label{tab:composition}\small{{\bf Properties of the modeled planet-impactor
				systems, and comparison with the observations of Solar system bodies.}
			${\rm M_{{\rm P}}}$,${\rm N_{{\rm P}}}$ and ${\rm M_{{\rm I}}}$,${\rm N_{{\rm I}}}$
			are the mass and number of particles in the planet and the impactor,
			respectively; $t_{{\rm coll}}$ is the collision time in the simulations;
			${\rm C_{{\rm cal}}}$is the calibration pre-factor (see main text).
			The $\Delta^{17}{\rm O}$ composition difference and the KS-probability
			(for the planet and impactor feeding-zone distribution to be sampled
			from the same parent distribution) are shown both the case of contribution
			of planetary material to the newly-formed Moon, and the cases of $20\%$ and $40\%$
			contribution of material from the planet.}}
\end{sidewaystable}

\newpage
\begin{figure}
	\center
	\includegraphics[scale=0.5]{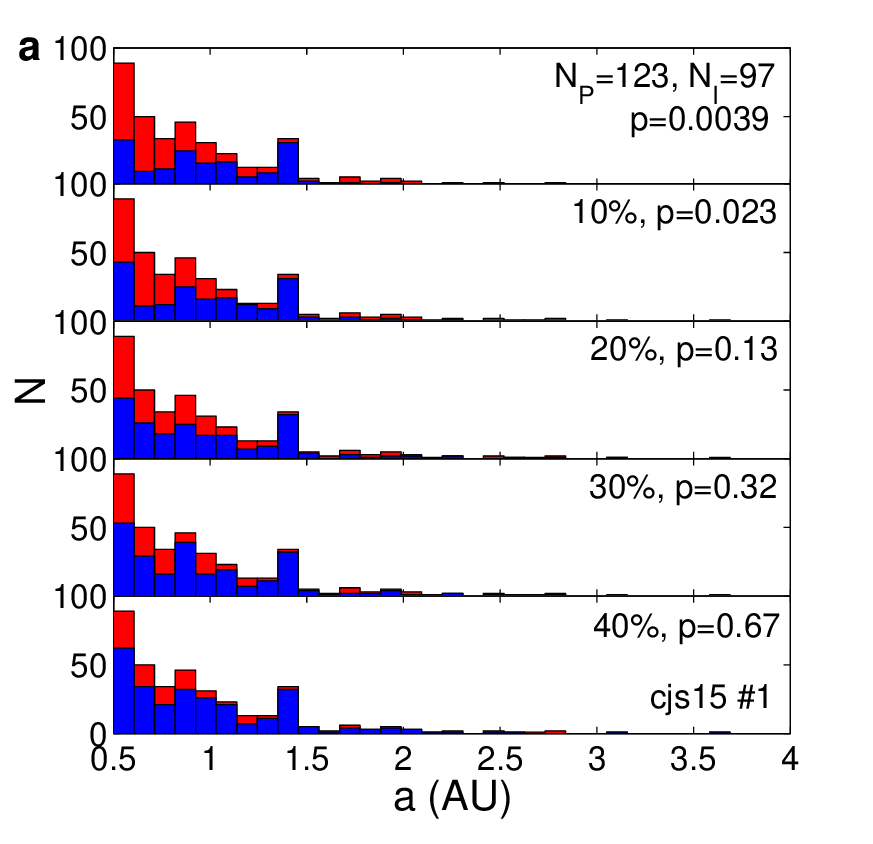}\includegraphics[scale=0.5]{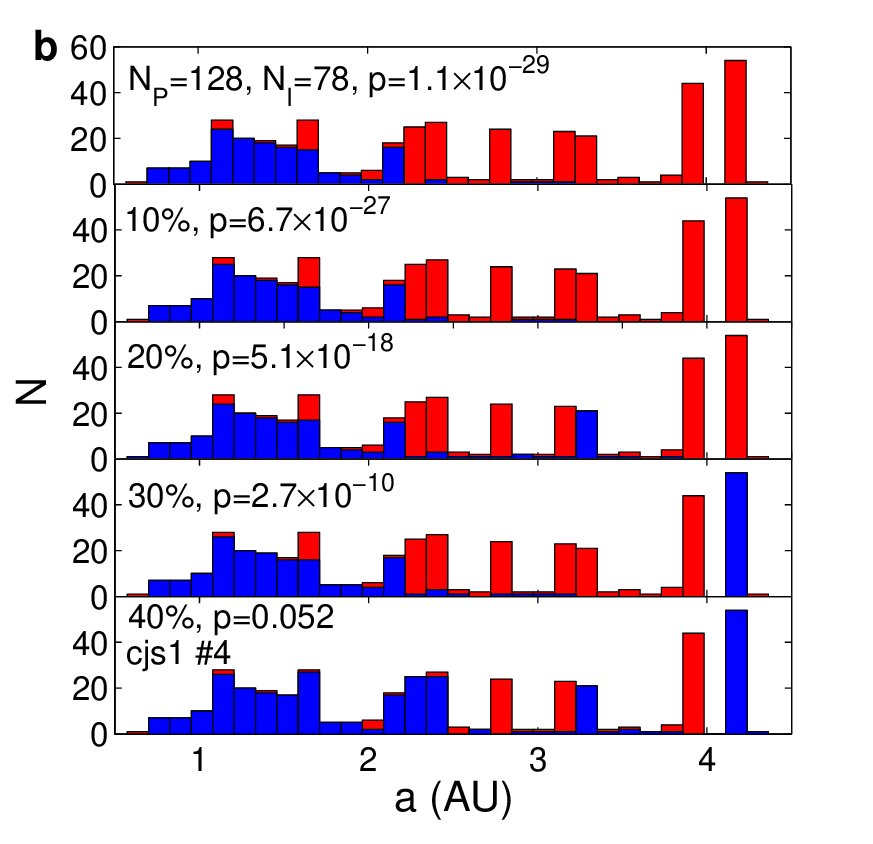}
	\caption{{\bf The distribution of planetesimals composing the planet
			and the impactor.} Panel {\bf{a}} shows a case where the
		origins of the planetesimals composing the planet and the impactor 
		are consistent with being sampled from
		the same parent distribution for the expected typical $20\%$ contribution of planetary material in moon-forming impacts (KS-test probability $>0.05$). 
		Panel {\bf{b}} shows a case where the planet and impactor compositions 
		are inconsistent ($P<0.05$), but become consistent once a significant ($40\%$) contribution of material from the planet is considered. The lower plots in each panel show the results assuming a different contributions from the planet  (four cases are shown $10\%,\,20\%,\,30\%$ and $40\%$). The cumulative distribution for these cases as well as all other planet-impactor pairs in Table 1 can be found in the Methods.}
\end{figure}

\begin{figure}
	\center
	\includegraphics[scale=0.33, trim=1cm 0cm 2.5cm 0.5cm, clip=true]{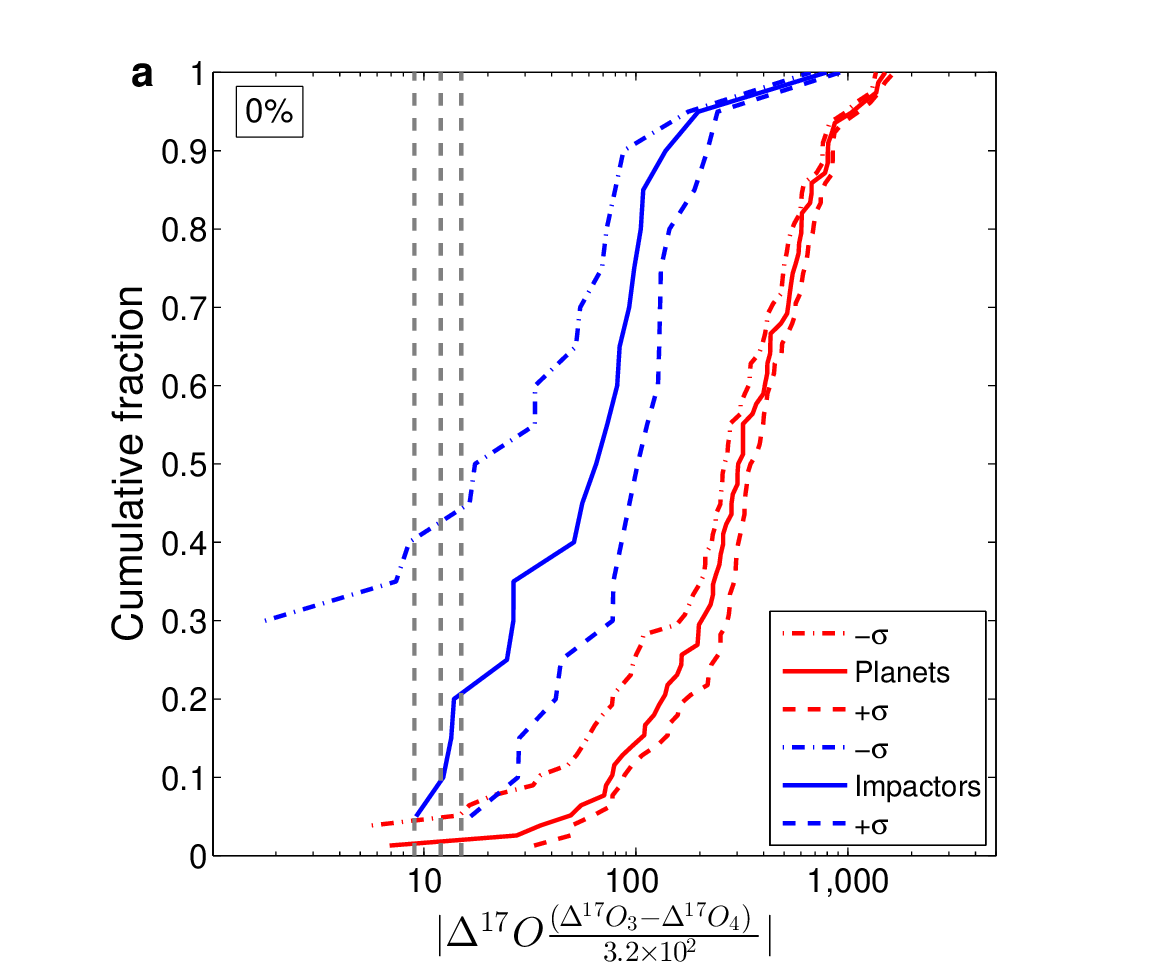}\includegraphics[scale=0.33, trim=1cm 0cm 2.5cm 0.5cm, clip=true]{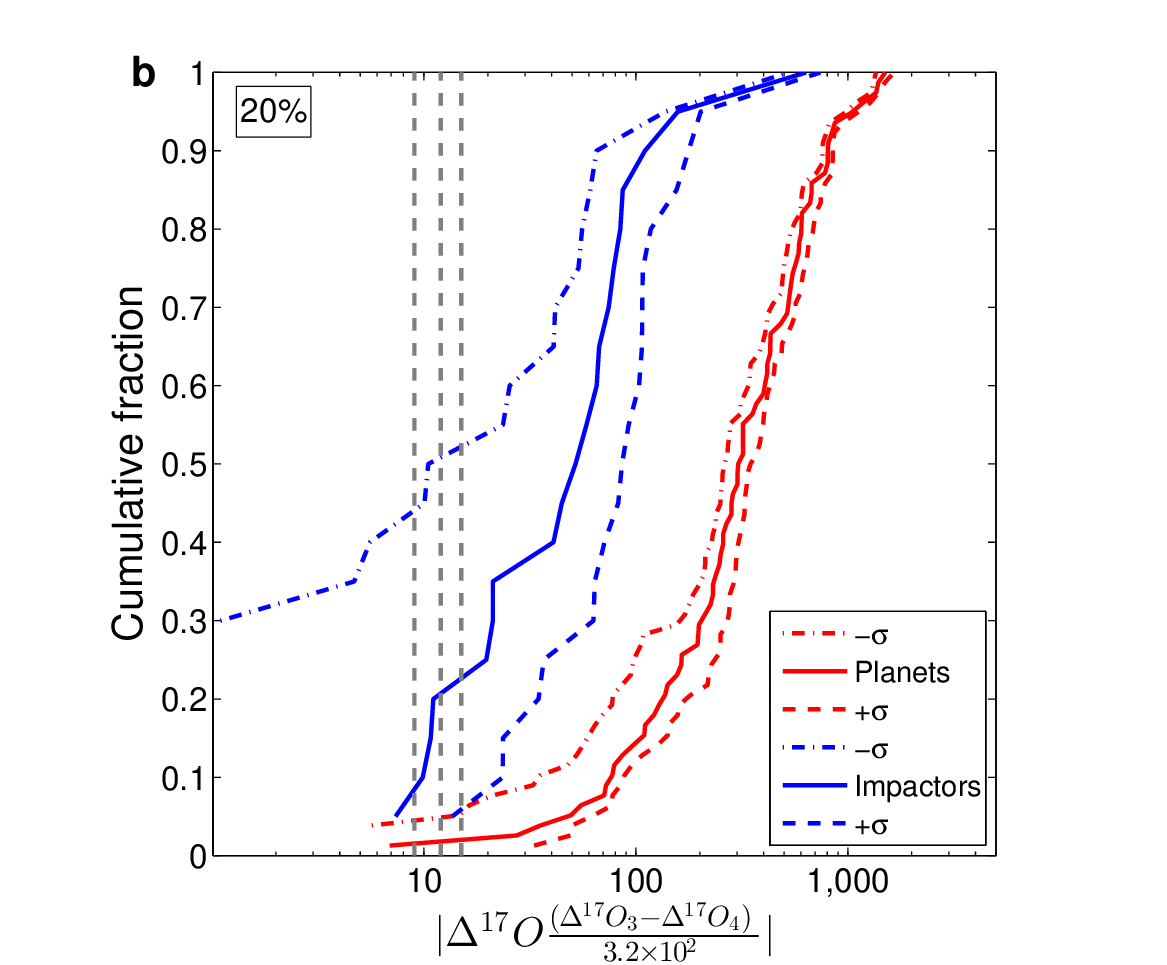}\includegraphics[scale=0.33, trim=1cm 0cm 2.5cm 0.5cm, clip=true]{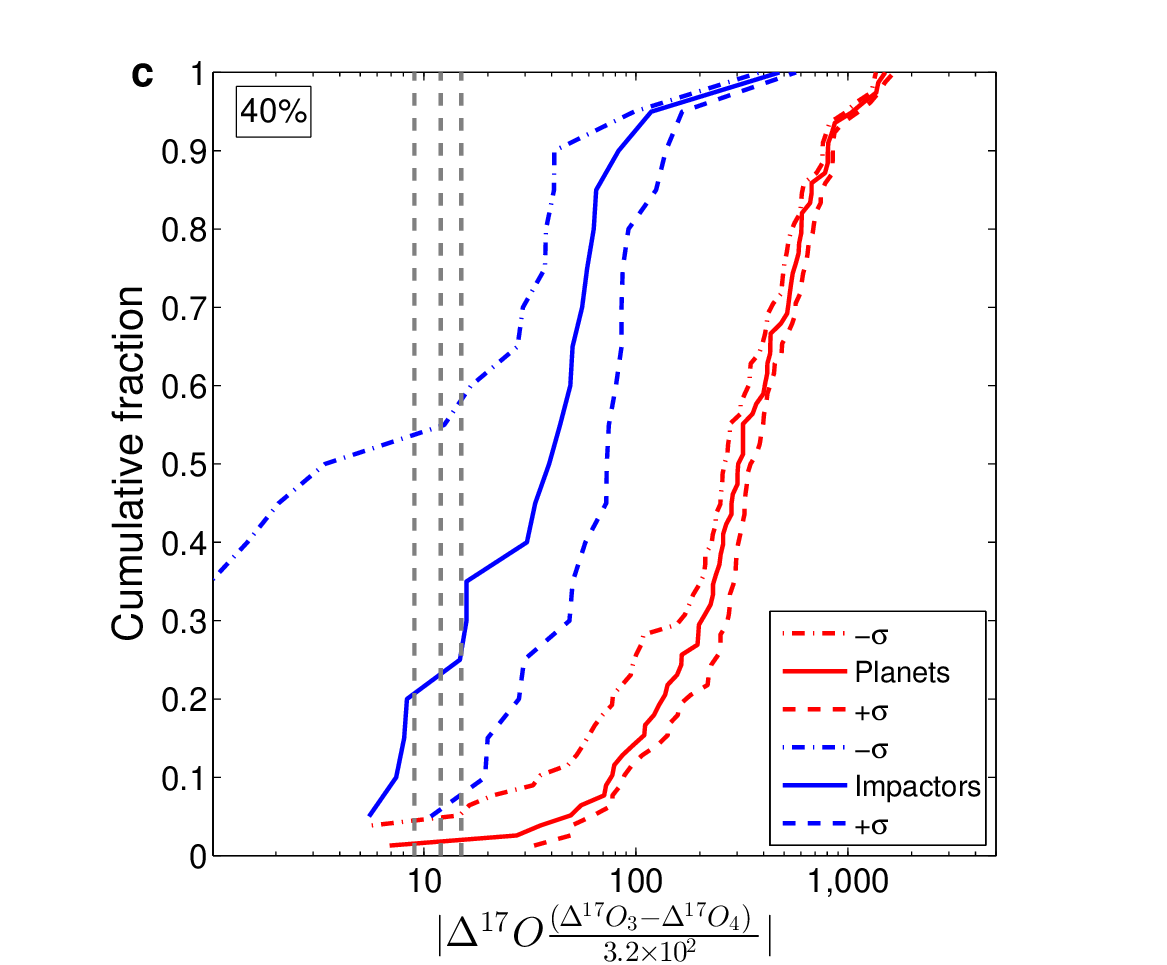}
	
	\caption{{\bf The cumulative distribution of the absolute $\Delta^{17}{\rm O}$ differences between planets and their last giant impactors (blue), compared with the differences between planets in the same system (red)}. Panels {\bf a}, {\bf b} and {\bf c} correspond to the cases of zero, $20\%$  and $40\%$ contribution of material from the planet to a Moon formed from these impacts, respectively. The vertical lines depict the $\Delta^{17}{\rm O}$ difference of the Earth-Moon system (dashed lines for the $\pm\sigma$). The differences between the planet-impactor pairs are systematically smaller than those found between different planets (the same parent distribution for the two groups can be excluded with high confidence; KS probability $6.7\times10^{-8}$,  $1.1\times10^{-8}$  and $1.3\times10^{-9}$ for the zero, $20\%$  and $40\%$ cases, respectively).}
\end{figure}
\clearpage

\part*{Methods}

In the following we supply additional data on the planet-impactor pairs' composition and the compositions of different planets at the same systems. We also provide a more detailed information on the methods used  as well as discuss the sensitivity of our results to the various criteria and calibrations which we applied. The full simulations data used in this work can be provided by the authors upon request.
 
\section*{Initial configuration of Jupiter and Saturn}
 The simulations analyzed here are described in detail in ref. \cite{Ra09}. These various simulations explore a range of different initial conditions for the gaseous planets.  In particular, in the cjs and cjsecc simulations Jupiter and Saturn are placed on  orbits with semimajor axes of 5.45 and 8.18 AU and mutual inclination of $0.5^\circ$. In cjs the orbits are circular while in cjsecc they are eccentric with $e_J=0.02$ and $e_S=0.03$. In eejs Jupiter and Saturn are placed on their current position (5.25 and 9.54 AU), with mutual inclination of $1.5^\circ$ and larger eccentricities  than the observed ones ($e_J=e_S=0.1$ or $e_J=0.07$ and $e_S=0.08$). In ejs the orbits of Jupiter and Saturn have similar parameters to those observed ($a_J=5.25$AU and $e_J=0.05$, $a_S=9.54$ and $e_S=0.06$) with mutual inclination of $1.5^\circ$. Finally, in jsres, Jupiter and Saturn are placed at $a_J=5.43$AU and  $a_S=7.30$AU with $e_J=0.07$ and $e_S=0.01$ and with mutual inclination of $0.2^\circ$.

\section*{The composition difference between planets in the same system}
The Supplementary Information Table shows the $\Delta^{17}{\rm O}$ differences between the different planets (in each system) and the impacted-planets analyzed in the main text. The full cumulative distribution for these data can be seen in Fig. 2 in the main text. Note that in those cases where two impacted planets were analyzed in the same system, the differences are shown both in respect to the first and second planets. 

\section*{The cumulative composition distribution of planet-impactor pairs}

We used the following procedure to calculate the spatial distribution of the feeding zones (used for  Fig. 1, Extended Data Fig. 1 and 2 and Table 1 KS-probabilities). We extracted the record of planetesimals that constitute the planet and the impactor before the last moon-forming impact, as well as the planet composition after the collision. In order to account for the different contribution coming from particles of different masses we replicated $n_i$ times each particles, where $n_i$ is the ratio between the mass of the $i$-Th. particle and the minimum mass of the planetesimals. The planetesimal record was then used to produce the distribution of the feeding zones used in our analysis. 
In cases where contribution of planetary material to the Moon composition was considered, we randomly chose particles from the planet and added them to the planetesimals composing the impactor, where appropriate numbers of particles were taken so as to produce the relevant fractional contribution (for the different cases of $10,\,20,\,30$ or $40\%$ contribution).  We then repeated the same analysis as done for the impactor, with these new mixed impactors. 

\section*{$\Delta^{17}O$ calibration }

In order to calculate the $\Delta^{17}{\rm O}(\text{\ensuremath{\equiv}}\delta^{17}O-0.52\delta^{16}O)$
for the planet and impactor pairs we followed the procedure described
by ref. \cite{PS07}. In order to assign specific values of $\Delta^{17}{\rm O}$
to each particle in the simulation we calibrated our simulations with
the solar system observations. We assume a linear gradient of $\Delta^{17}{\rm O}$
with heliocentric distance $r$

\begin{equation}
\text{\ensuremath{\Delta}}^{17}{\rm O}(r)=c_{1}r+c_{2}.
\end{equation}
where the two free parameters in Equation 1 were calibrated imposing
that the third planet formed in the system has the composition of
the Earth ($\Delta^{17}{\rm O}=0\permil$) and the fourth one has the composition
of Mars ($\Delta^{17}{\rm O}=+0.32\permil$). In cases where only
three planets formed (marked with a $^{*}$ in Table 1), we assigned
the composition of the Earth and Mars to the second and third planets
in the simulation, respectively. We then  mass-averaged over all the 
$\text{\ensuremath{\Delta}}^{17}{\rm O}(r)$
of the planetesimals that accreted forming the Earth. We used the
initial position of each body as the heliocentric distance $r$. We
did the same for the planetesimals composing ``Mars''. In this way we
have the system of equations

\begin{align*}
\begin{cases}
\frac{\overset{N}{\sum}m_{i,E}(c_{1}r_{i,E}+c_{2})}{M_E} & =0\\
\frac{\overset{N'}{\sum}m_{i, M}(c_{1}r_{i,M}+c_{2})}{M_M} & =0.32,
\end{cases}\\\textbf{}
\end{align*}
 where $r_{i,E}$ and $m_{i,E}$  ($r_{i,M}$ and $m_{i,M}$) are the initial position and mass of the i$^{th}$
planetesimal composing the Earth (Mars) and $M_E$ ($M_M$) is the 
final total mass of the Earth (Mars). In this way it has
been possible to evaluate the $\Delta^{17}{\rm O}$ value for each planetesimal
in the system and thus for all the planets in each system. Given this
calibration we evaluate the $\text{\ensuremath{\Delta}}^{17}{\rm O}$ of
the planet ($\text{\ensuremath{\Delta}}^{17}{\rm O}_{P}$) and of the last
impactor ($\text{\ensuremath{\Delta}}^{17}{\rm O}_{I}$), as the average
of the $\text{\ensuremath{\Delta}}^{17}{\rm O}$ of all their respective
components, as well as calculated the $1\sigma$ SEM for each of these
values ($\sigma_{P}$ and $\sigma_{I}$). To check whether or not
the planet-Moon system is consistent with the Earth-Moon system we
evaluated the difference $\text{\ensuremath{\Delta}}^{17}{\rm O}_{I}-\text{\ensuremath{\Delta}}^{17}{\rm O}_{P}$
and the relative error $\sigma=\sqrt{\sigma_{P}^{2}+\sigma_{I}^{2}}$.

In order to calculate the $\Delta^{17}{\rm O}$ when a fractional contribution from the planet is included, we added the average  $\Delta^{17}{\rm O}$ of the planet and the impactor, each weighted according to the appropriate fractional contribution considered.

In order to study the sensitivity of our results to the Earth-Mars composition difference calibration used,  we also considered lower and higher calibrations, between 0.5-1.5 times the Earth-Mars  $\Delta^{17}{\rm O}$ difference. We re-analyzed the fraction of compatible planet-impactor pairs (i.e. producing planet-moon pairs with composition difference equal or smaller than the Earth-Moon composition difference) for these different calibrations. The results are summarized in Extended Data Table 1.  We find that although the fraction of consistent pairs decrease with the use of larger difference calibration, as expected, difference calibrations as much as 1.5 larger than the Earth-Mars difference still give rise to a mean 5\% of planet-impactor pairs with similar composition (and 40\% within 1$\sigma$ uncertainty in the simulation compositions), rising to a mean 10\% to 20\% for the cases of 20\% and 40\% mixing of the planetary material, respectively. In other words, the results are generally robust to that level and do not dependent on a fine-tuned calibration.  

\section*{Dependence on the criteria for the planet-impactor pairs considered in the analysis}
 In order to verify the robustness of our results to different criteria for the choice of planet-impactor pairs used in our analysis we studied various different criteria:  
(1) Use all planet - last-impactor pairs, considering smaller thresholds (i.e. smaller than the 50 particles threshold considered in the main text) for the number of composing particles (and corresponding masses), but requiring an impactor mass of at least $0.5$ M$_{\rm Mars}$ to assure a moon-forming impact. Taking a threshold of 1, 10, 20, 30 and 40 minimal number of particles, we find that the general conclusion is unchanged; impactors have more similar composition to the planets they impact compared with other planets in the system. The mean fraction of planet-impactor pairs with comparable similarity as the Earth-Moon system is between $\approx10-20$\% in all cases (and up to $20-40$\% considering the 1 $\sigma$ uncertainties), as shown in Extended Data Table 2. Extended Data Figures 3, 4 and 5 show the cumulative distributions of the composition of planets and last impactors for all the systems, regardless of the number of particles that contributed to their formation, and for minimum of 10, 20, 40 and 50 particles composing the planet and 
last impactor. The Extended Data Figures 3, 4 and 5 are shown for 0\%, 20\% and 40\% mixing between planet's and impactor's material.\\
(2) Consider only last-impactors on the third planet. Once we require the impactor to have at least $0.5$ M$_{\rm Mars}$ and be composed of a significant number of particles the statistics become too small. The only case for which we have a significant (18 cases) is when we do not consider a minimal threshold for the number of composing particles. In this case we find 2 out of 18 (11\%; up to $\approx30$ \% when with the 1 $\sigma$ uncertainty in the simulations composition) planet-impactor pairs with composition difference equal or smaller than the Earth-Moon system.

\setcounter{table}{0}
\begin{table}[H]
\center\center\def\tablename{Extended Data Table}
\begin{tabular}{|c|c|c|c|c|c|c|}
\hline 
\multirow{2}{*}{Factor} & \multicolumn{2}{c|}{$0$\%} & \multicolumn{2}{c|}{$20$\%} & \multicolumn{2}{c|}{$40$\%}\tabularnewline
\cline{2-7} 
 & Mean & $1\sigma$ & Mean & $1\sigma$ & Mean & $1\sigma$\tabularnewline
\hline 
\hline 
$0.5$ & $35\%$ & $50\%$ & $35\%$ & $60\%$ & $35\%$ & $70\%$\tabularnewline
\hline 
$0.75$ & $20\%$ & $50\%$ & $25\%$ & $50\%$ & $35\%$ & $60\%$\tabularnewline
\hline 
$1$ & $20\%$ & $40\%$ & $20\%$ & $50\%$ & $25\%$ & $55\%$\tabularnewline
\hline 
$1.25$ & $5\%$ & $40\%$ & $20\%$ & $50\%$ & $20\%$ & $50\%$\tabularnewline
\hline 
$1.5$ & $5\%$ & $40\%$ & $10\%$ & $40\%$ & $20\%$ & $50\%$\tabularnewline
\hline 
\end{tabular}

\caption{{\bf The mean fraction of last impactors with compatible composition
respect to the planet they impact is given for different normalization
factors and mixing percentages.} The pairs which are consistent within $1\sigma$ of the simulation uncertainties are also given.}
\end{table}

\begin{table}
\center\def\tablename{Extended Data Table}
\begin{tabular}{|c|c|c|c|c|c|c|c|}
\hline 
\multirow{2}{*}{$N_{min}$} & \multicolumn{2}{c|}{$0$\%} & \multicolumn{2}{c|}{$20$\%} & \multicolumn{2}{c|}{$40$\%} & \multirow{2}{*}{$N_{cases}$}\tabularnewline
\cline{2-7} 
 & Mean & $1\sigma$ & Mean & $1\sigma$ & Mean & $1\sigma$ & \tabularnewline
\hline 
\hline 
$0$ & $10.1\%$ & $21.3\%$ & $10.1\%$ & $25.8\%$ & $11.2\%$ & $33.7\%$ & $89$\tabularnewline
\hline 
$10$ & $10.5\%$ & $24.6\%$ & $10.5\%$ & $29.8\%$ & $12.3\%$ & $36.9\%$ & $57$\tabularnewline
\hline 
$20$ & $12.5\%$ & $30\%$ & $12.5\%$ & $35\%$ & $15\%$ & $42.5\%$ & $40$\tabularnewline
\hline 
$30$ & $14.3\%$ & $35.7\%$ & $14.3\%$ & $42.9\%$ & $17.9\%$ & $46.4\%$ & $28$\tabularnewline
\hline 
$40$ & $18.2\%$ & $36.4\%$ & $18.2\%$ & $45.5\%$ & $22.7\%$ & $50\%$ & $22$\tabularnewline
\hline 
$50$ & $20\%$ & $40\%$ & $20\%$ & $50\%$ & $25\%$ & $55\%$ & $20$\tabularnewline
\hline 
\end{tabular}

\caption{{\bf 
The mean fraction of planet-impactor consistent pairs is shown for different mixing percentages
(0, 20, 40\%) and minimum numbers ($N_{min}$) of particles composing the impactor and planet (1, 10, 20,
30, 40, 50). }  The pairs which are consistent within $1\sigma$ of the simulation uncertainties are also given.
Only the $N_{cases}$ cases in which the last-impactor has a mass $>0.5$M$_{\rm Mars}$ have been taken into account. }
\end{table}
\clearpage

\setcounter{figure}{0}
\begin{figure}\center\def\figurename{Extended Data Figure}
\includegraphics[scale=0.25]{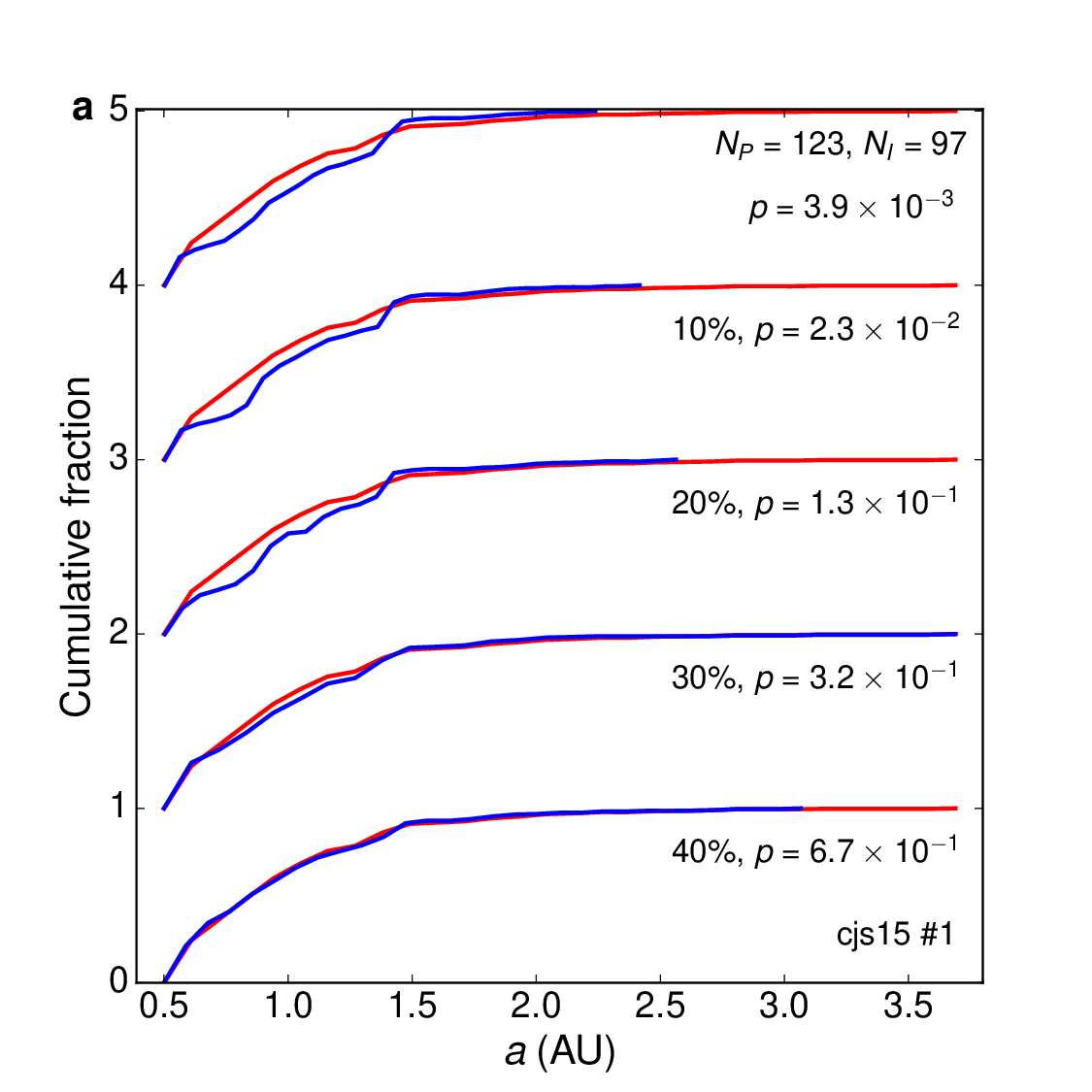}\includegraphics[scale=0.25]{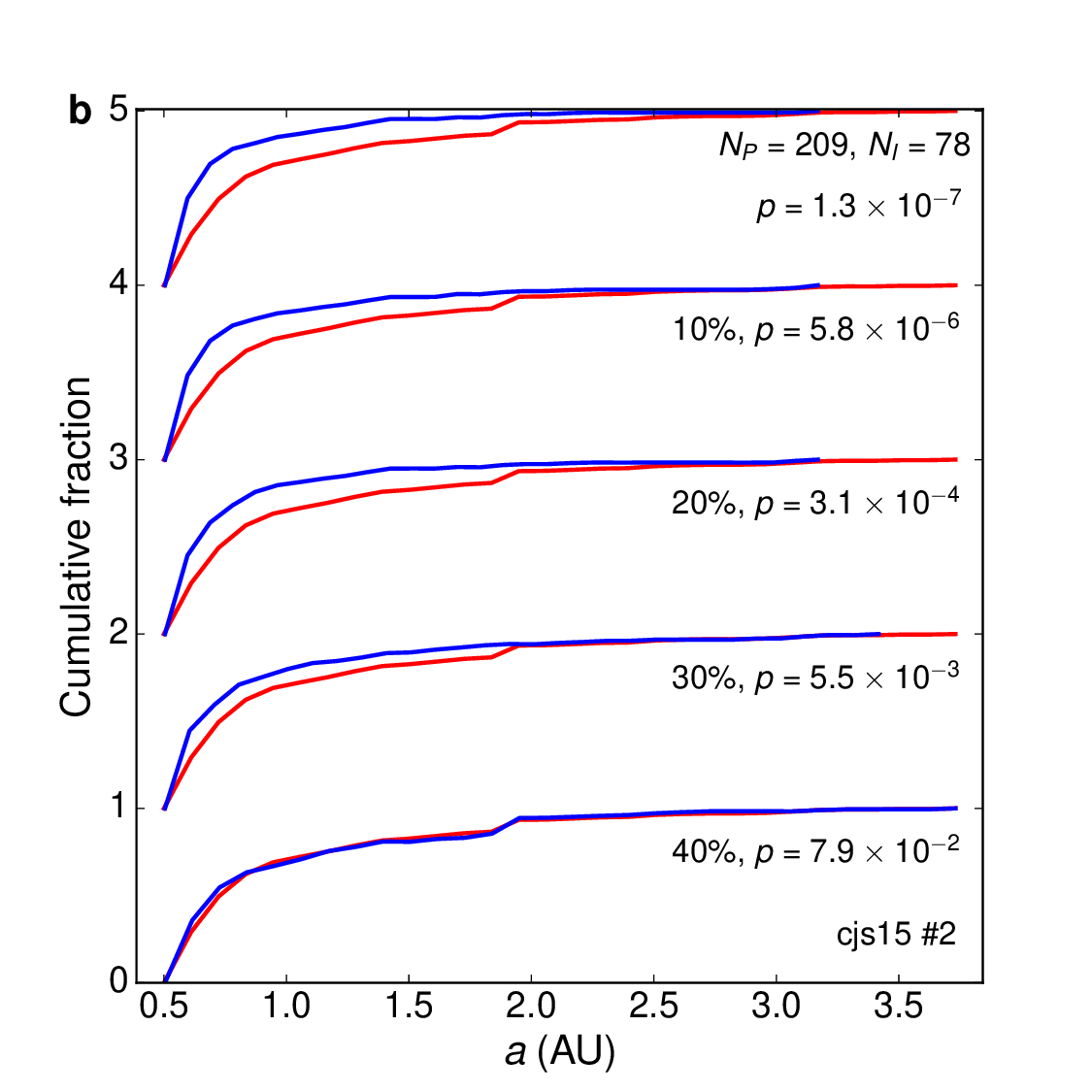}\includegraphics[scale=0.25]{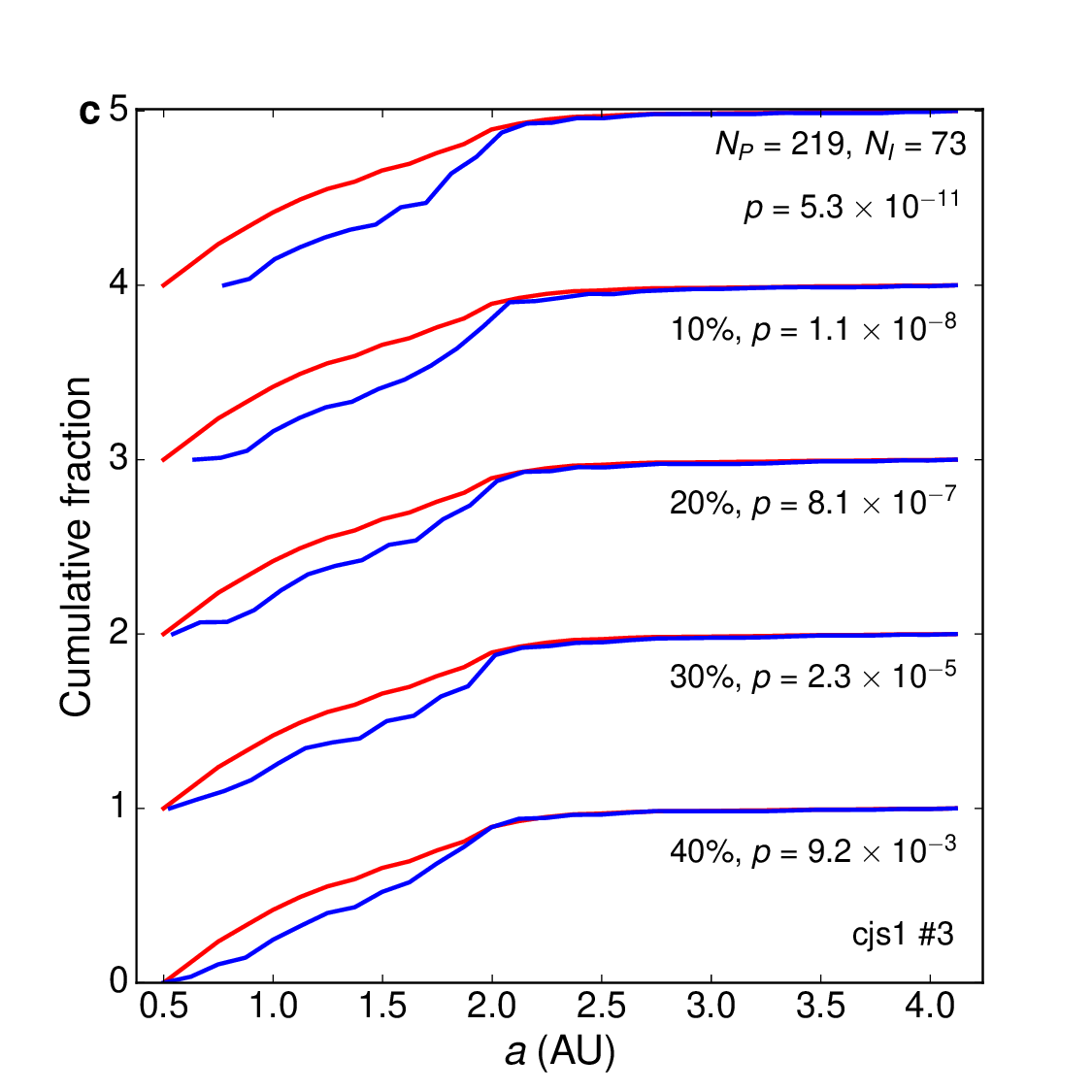}

\includegraphics[scale=0.25]{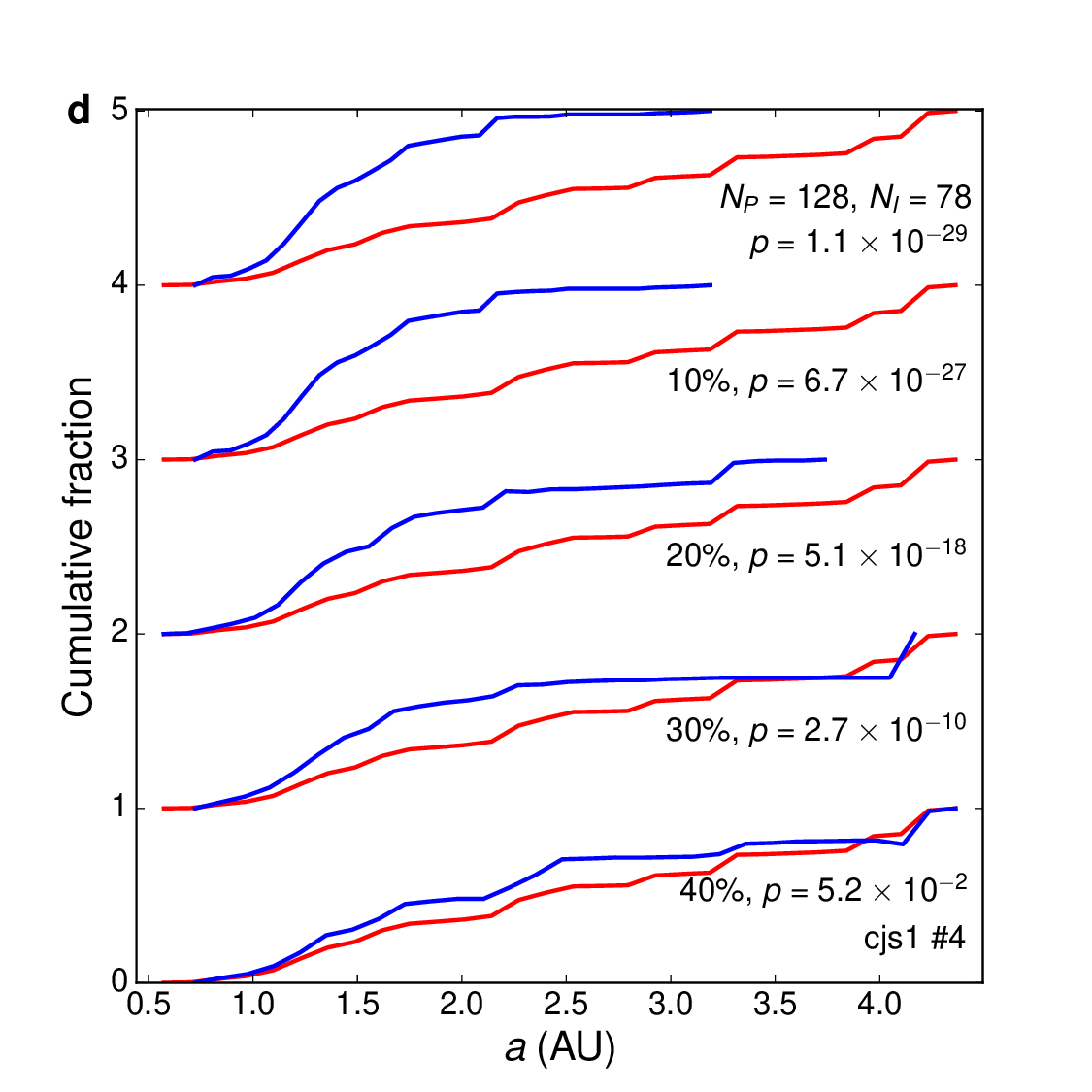}\includegraphics[scale=0.25]{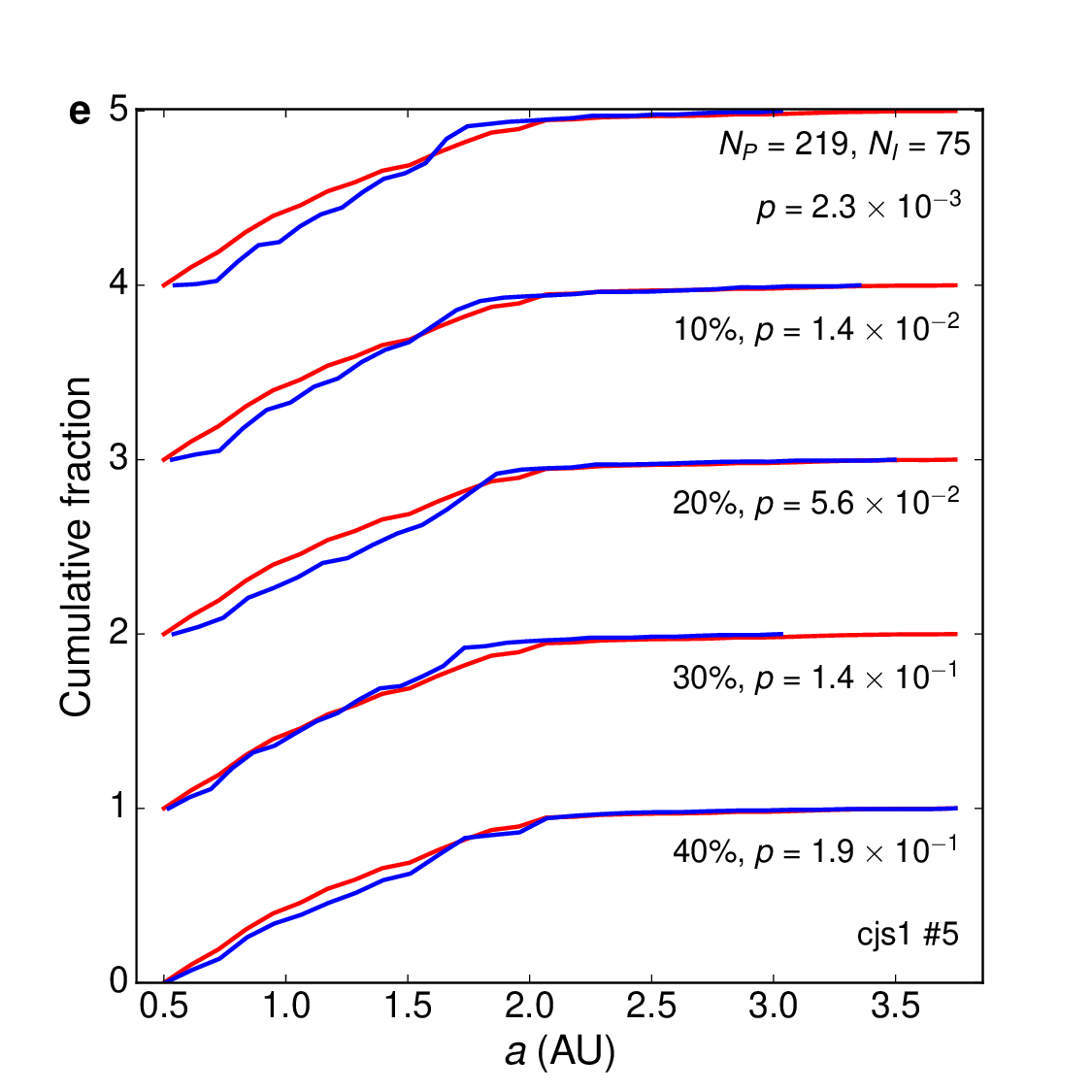}\includegraphics[scale=0.25]{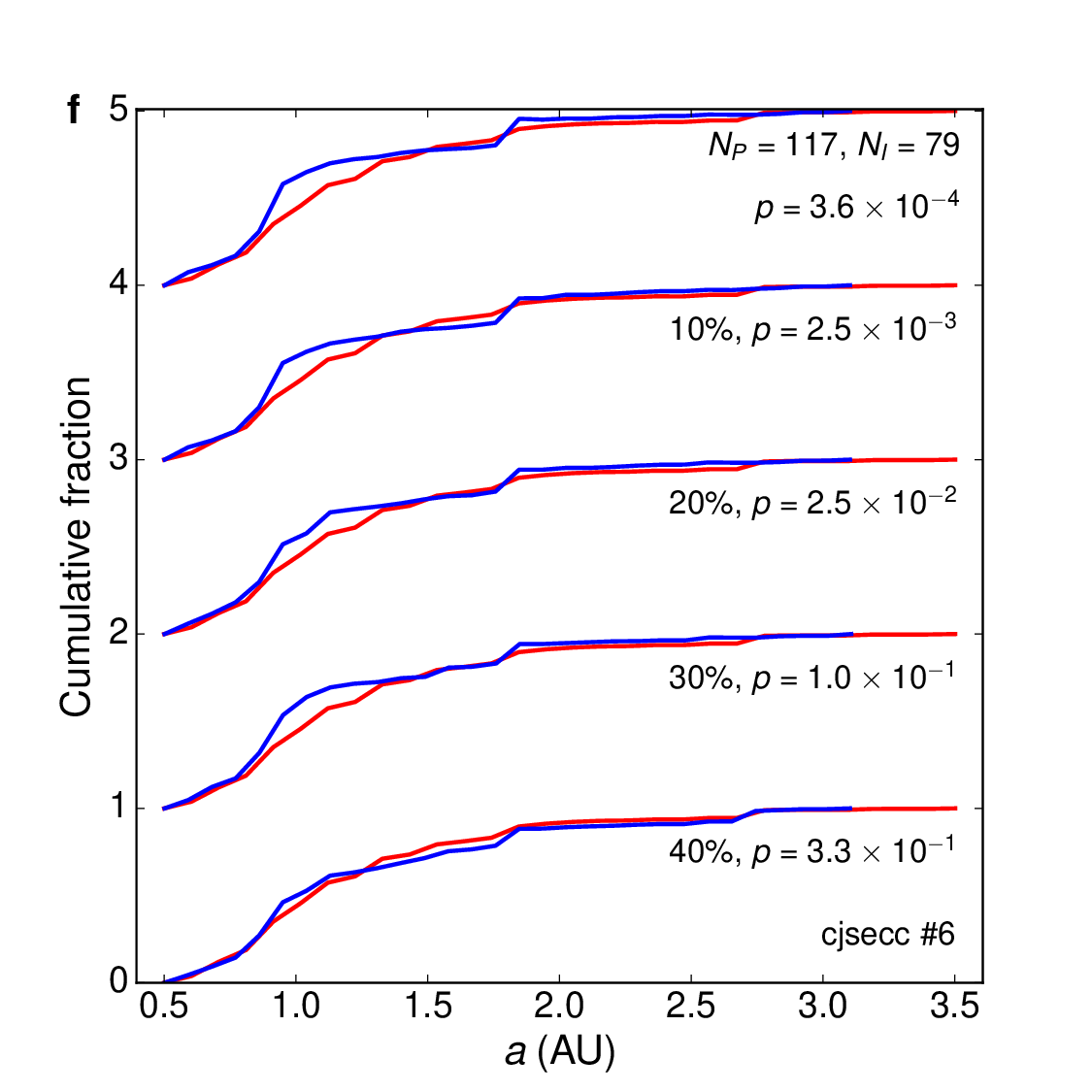}

\includegraphics[scale=0.25]{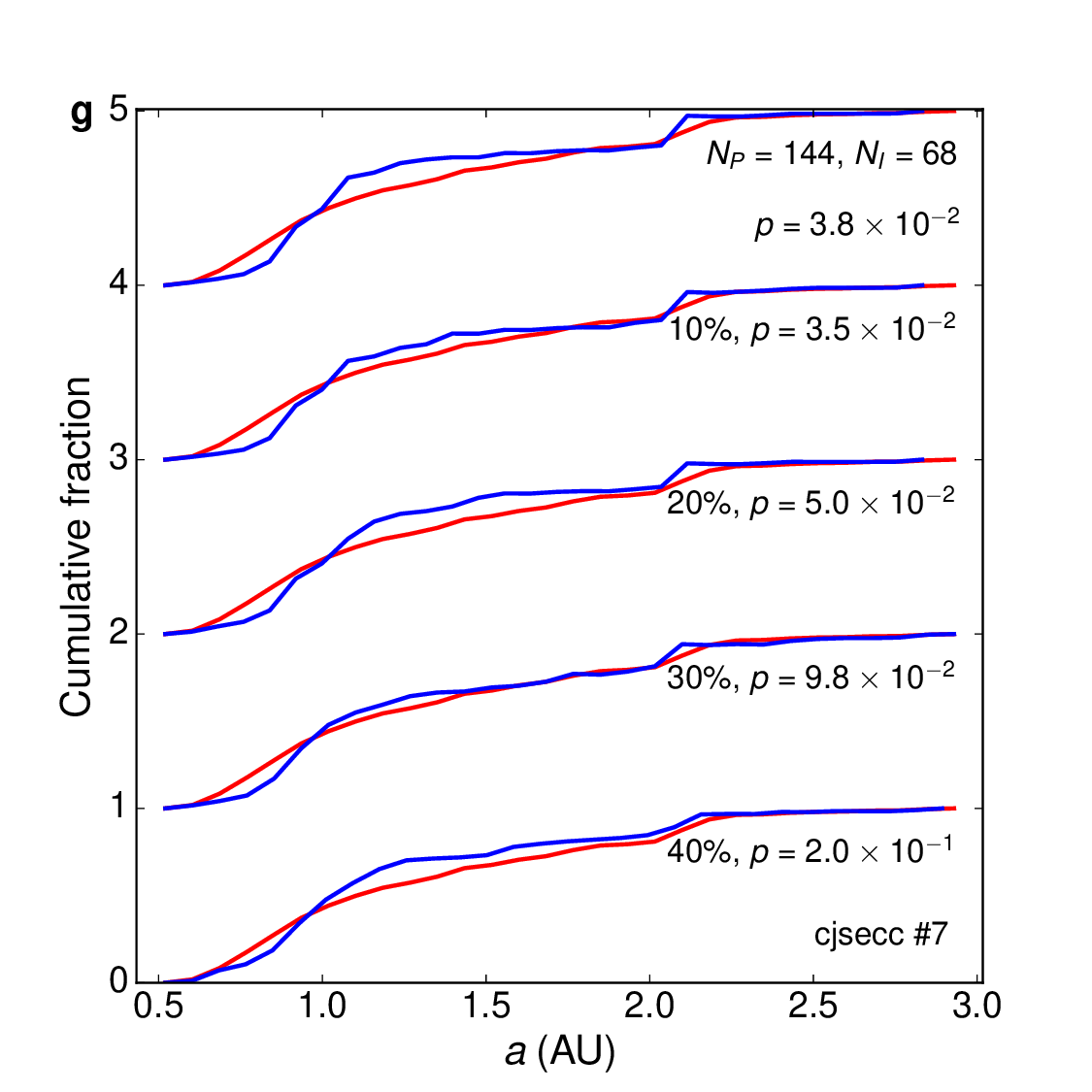}\includegraphics[scale=0.25]{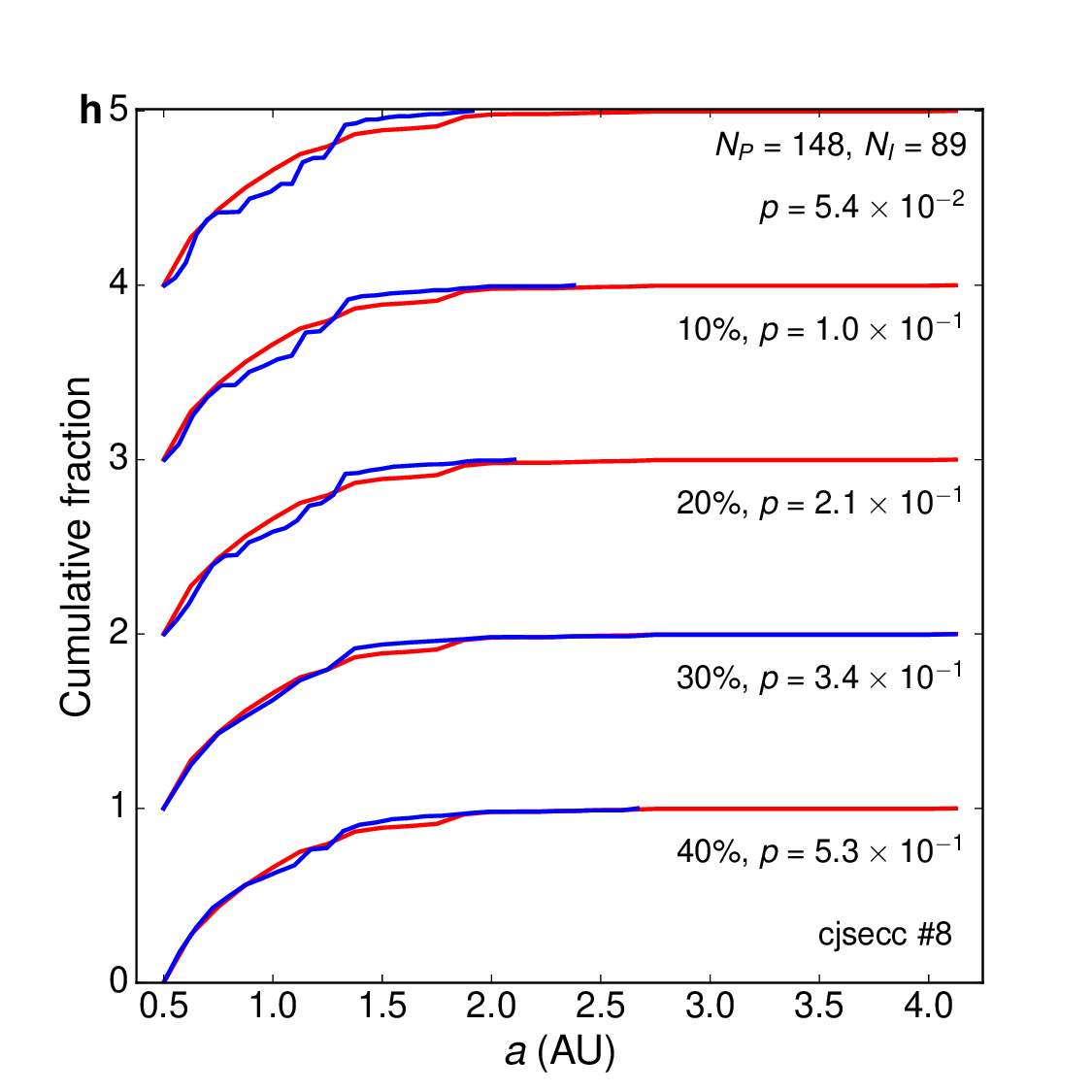}\includegraphics[scale=0.25]{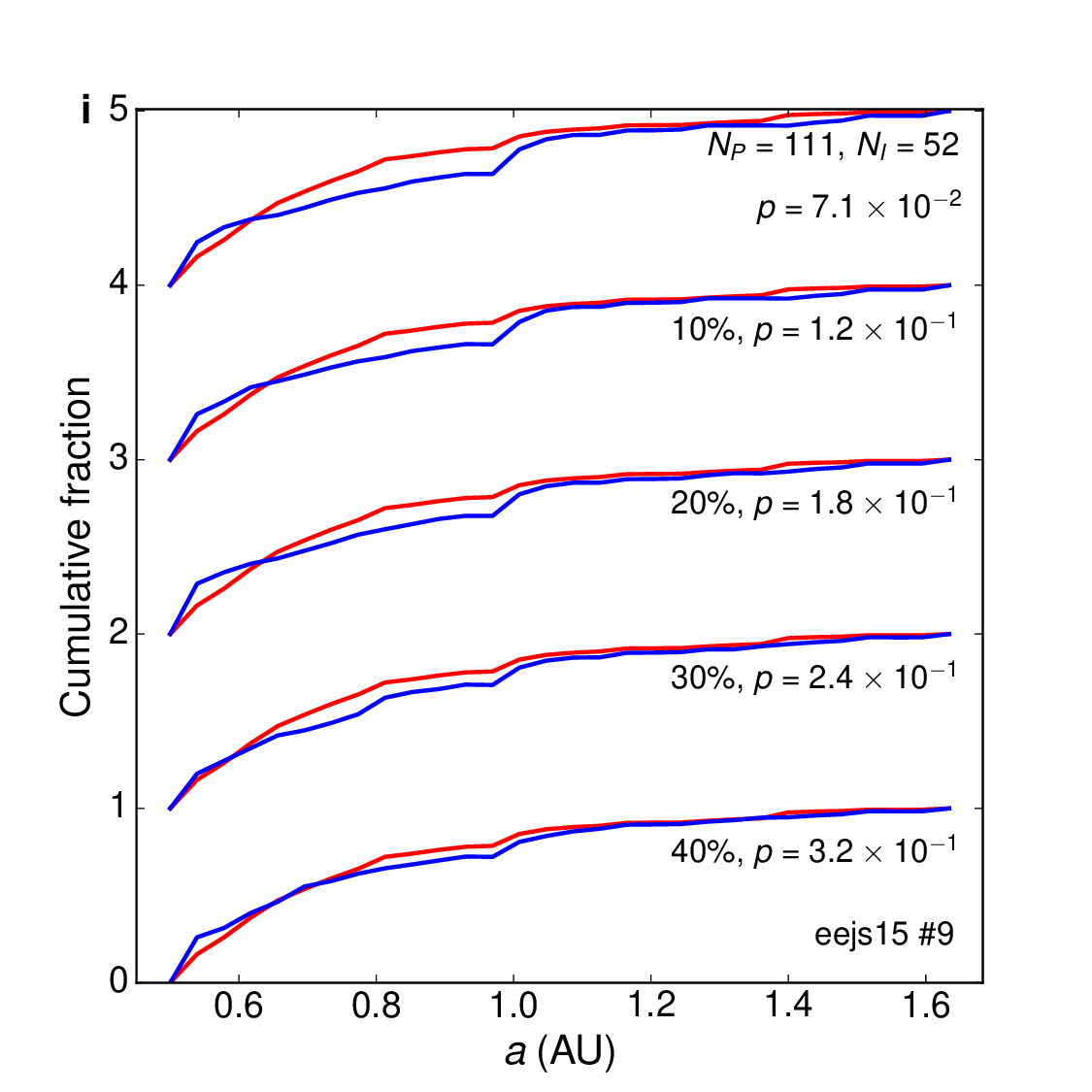}

\includegraphics[scale=0.25]{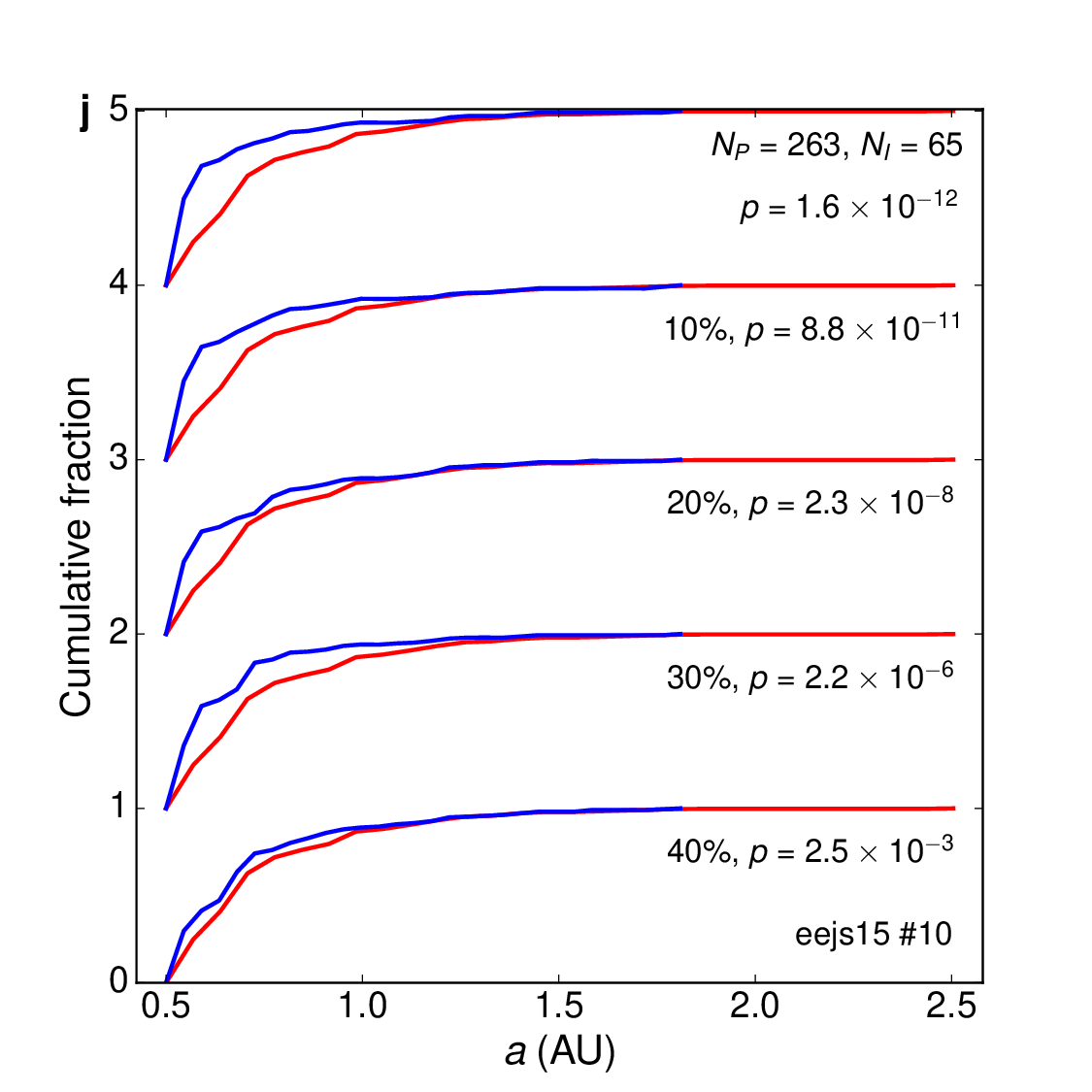}\includegraphics[scale=0.25]{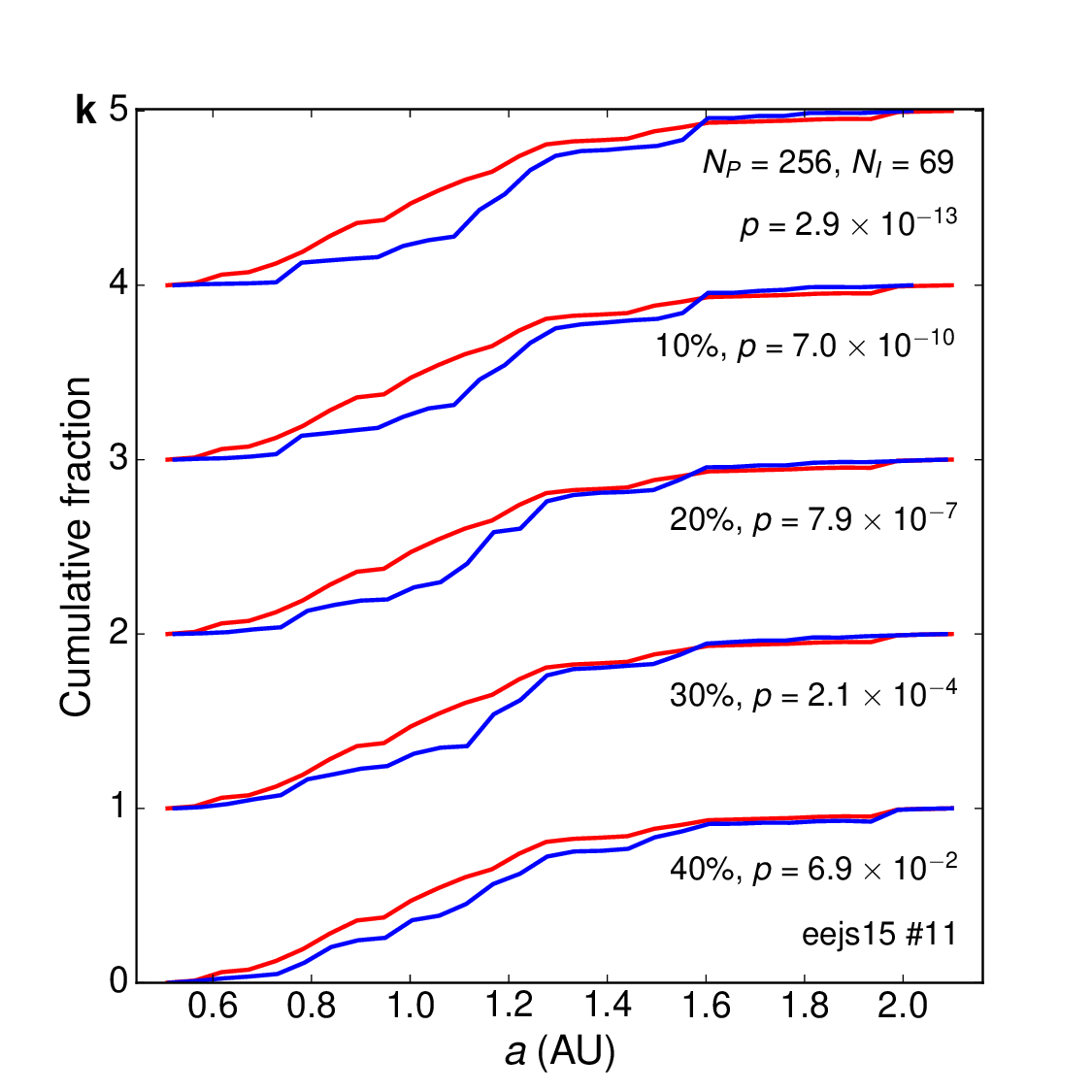}\includegraphics[scale=0.25]{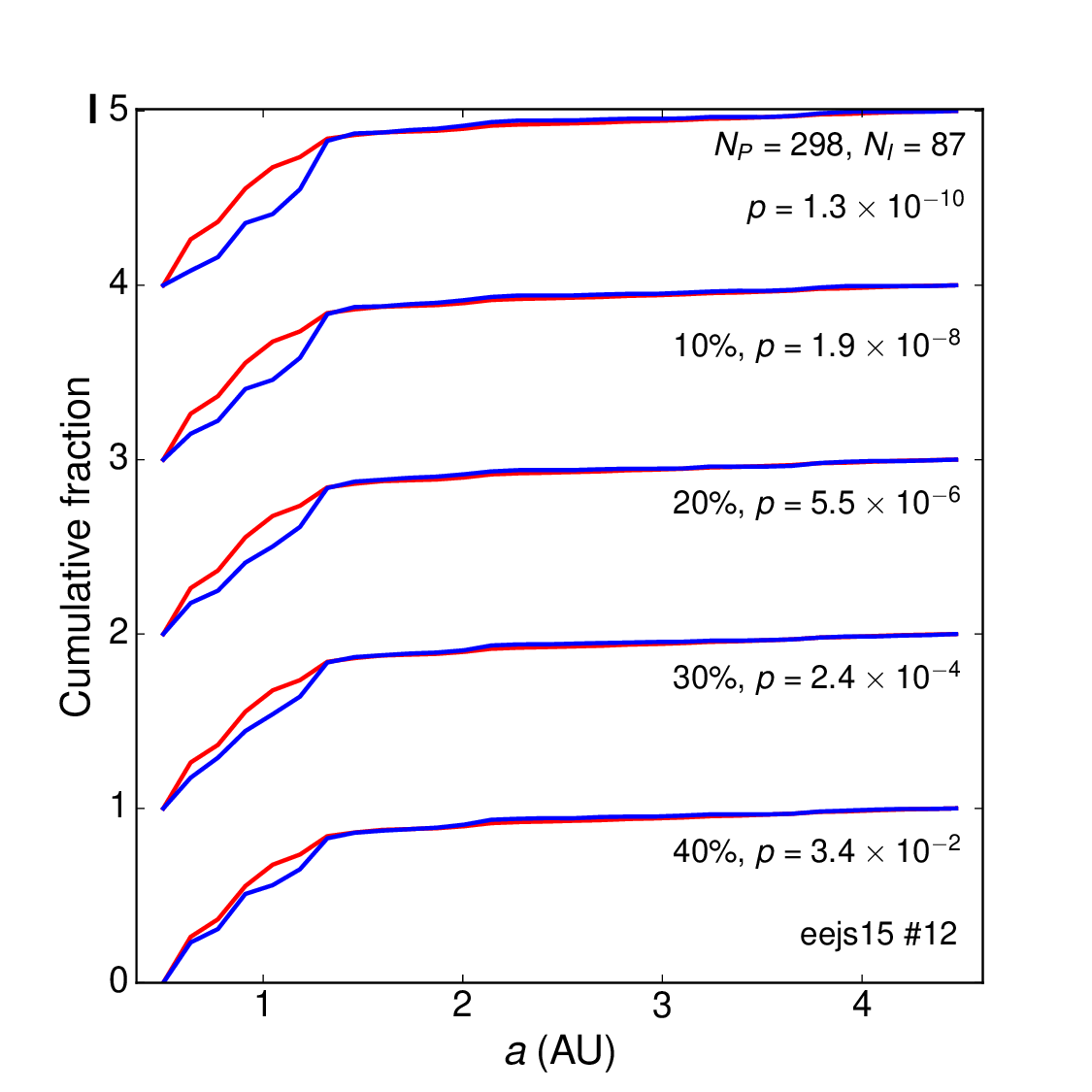}

\caption{{\bf The cumulative distribution of the planetesimals composing the planet
and the impactor}, showing all planet-impactor pairs in Table 1, cases 1-12 (panels {\bf a}-{\bf l}), including the cumulative distributions corresponding to the histograms in Fig. 1 in the main text.}
\end{figure}

\begin{figure}\center\center\def\figurename{Extended Data Figure}
\includegraphics[scale=0.25]{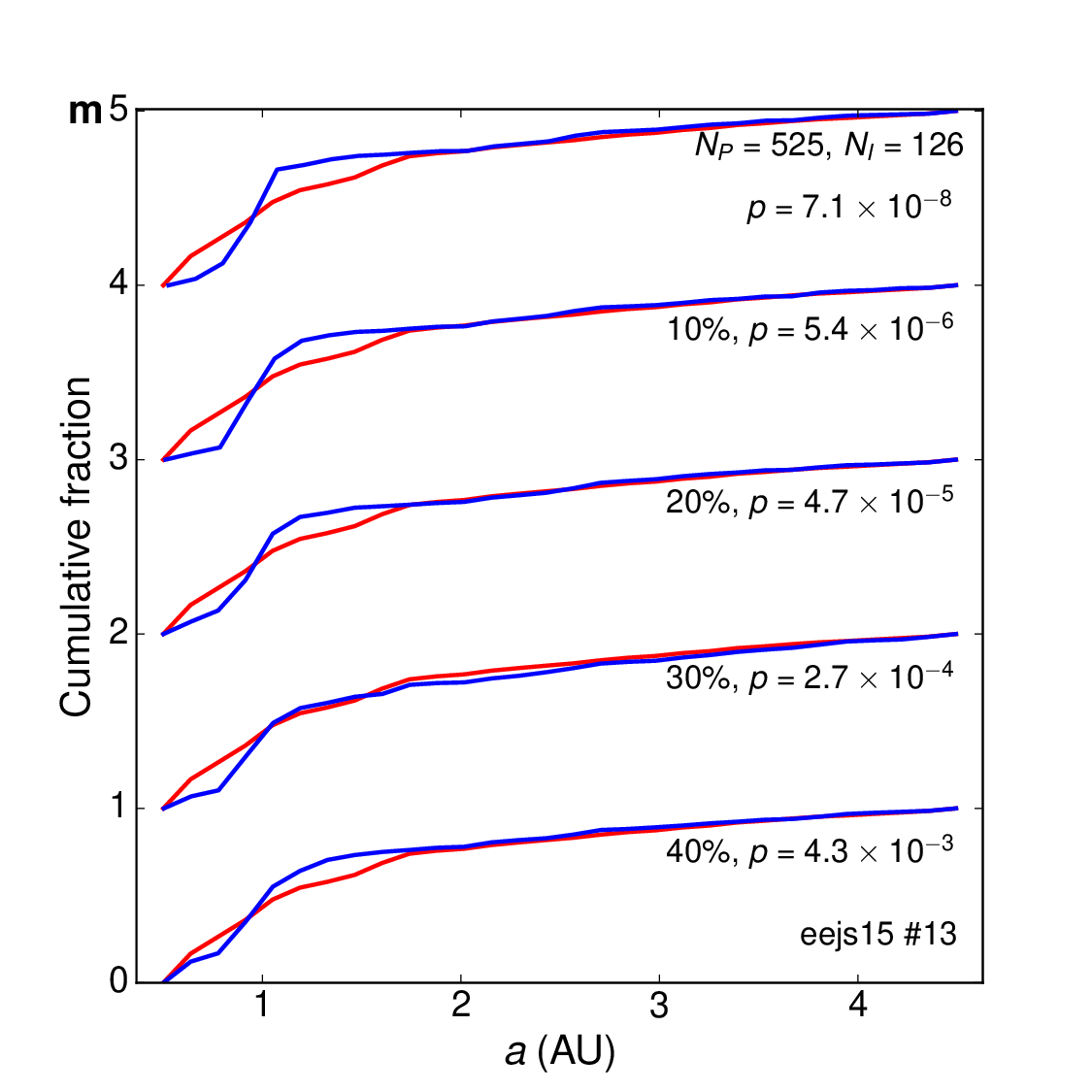}\includegraphics[scale=0.25]{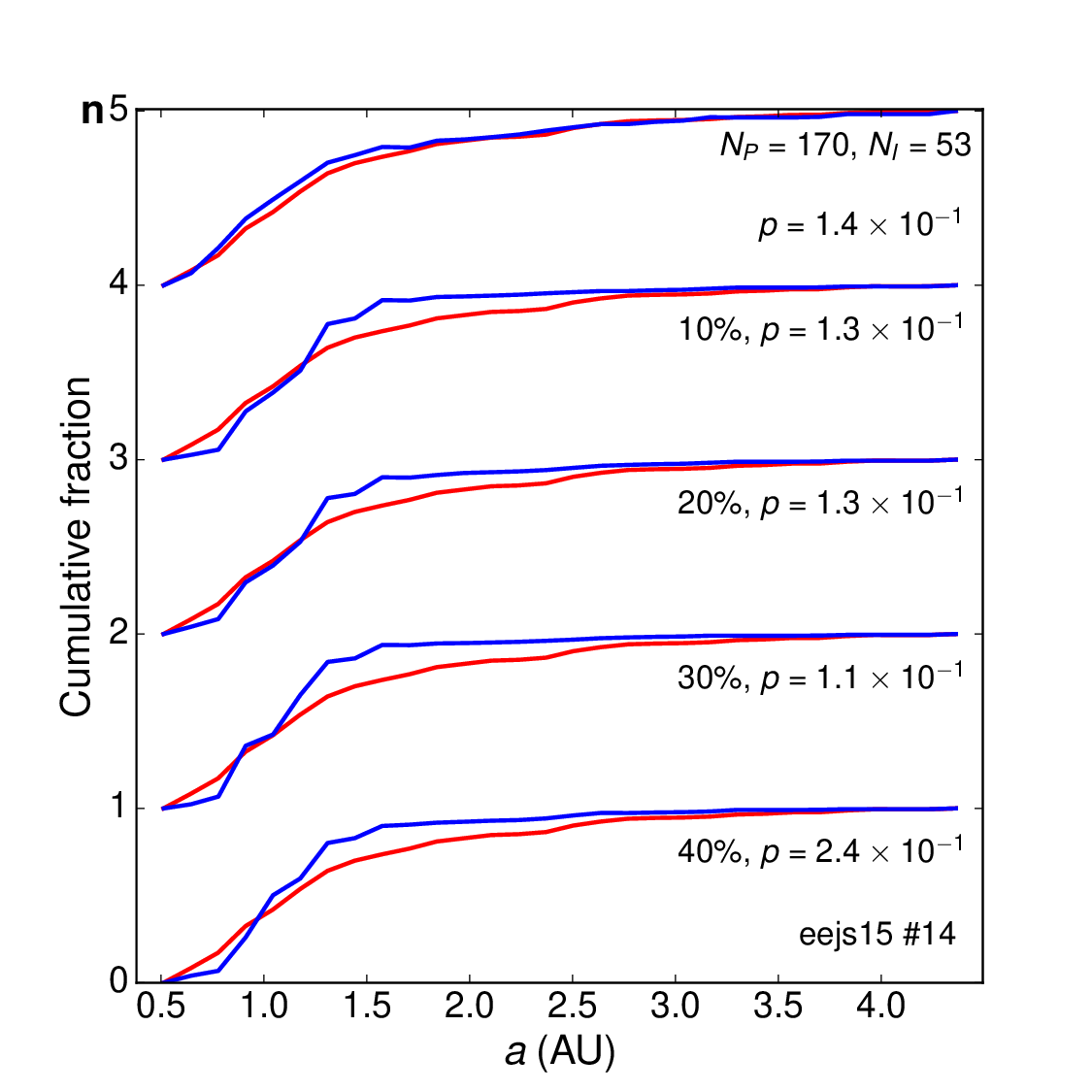}\includegraphics[scale=0.25]{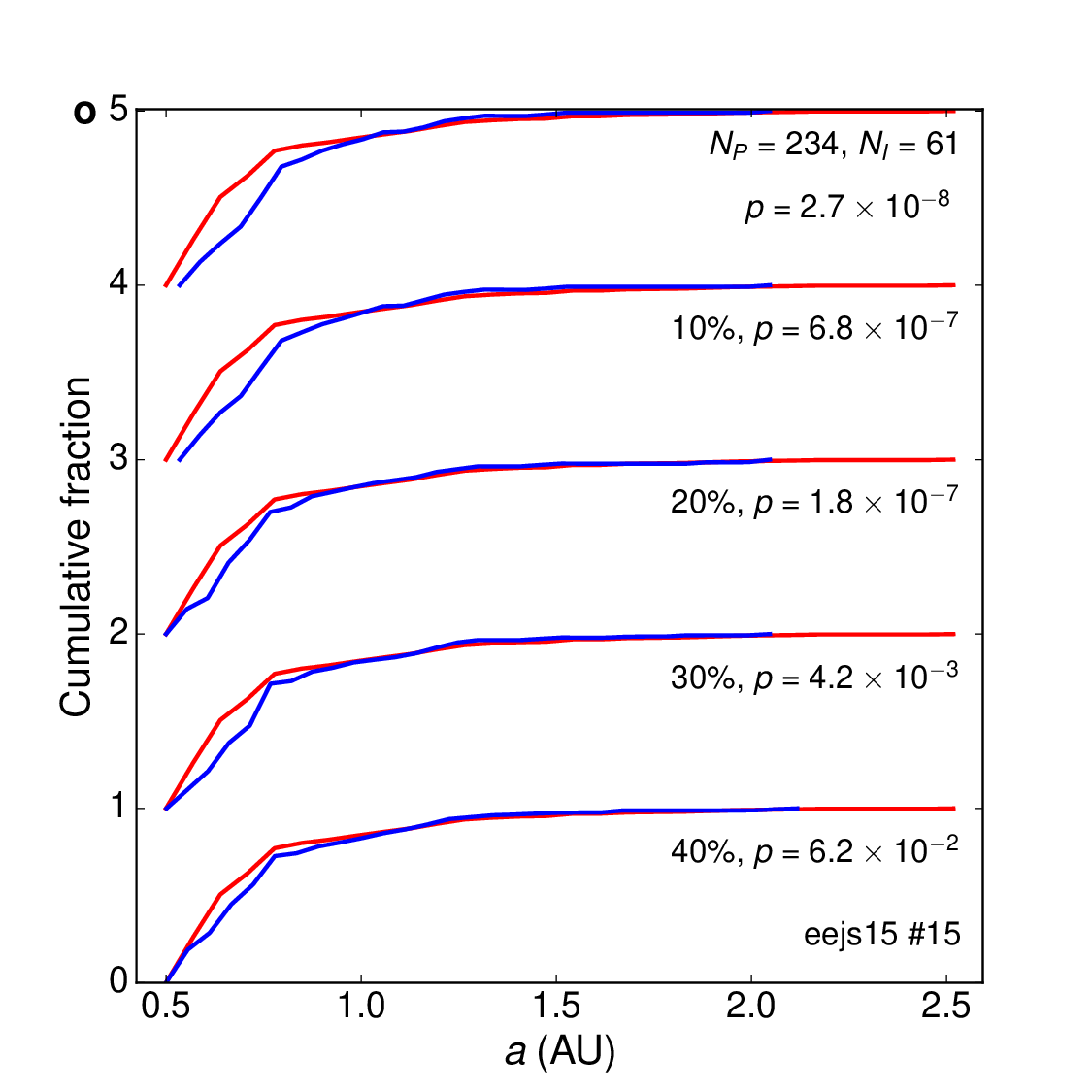}

\includegraphics[scale=0.25]{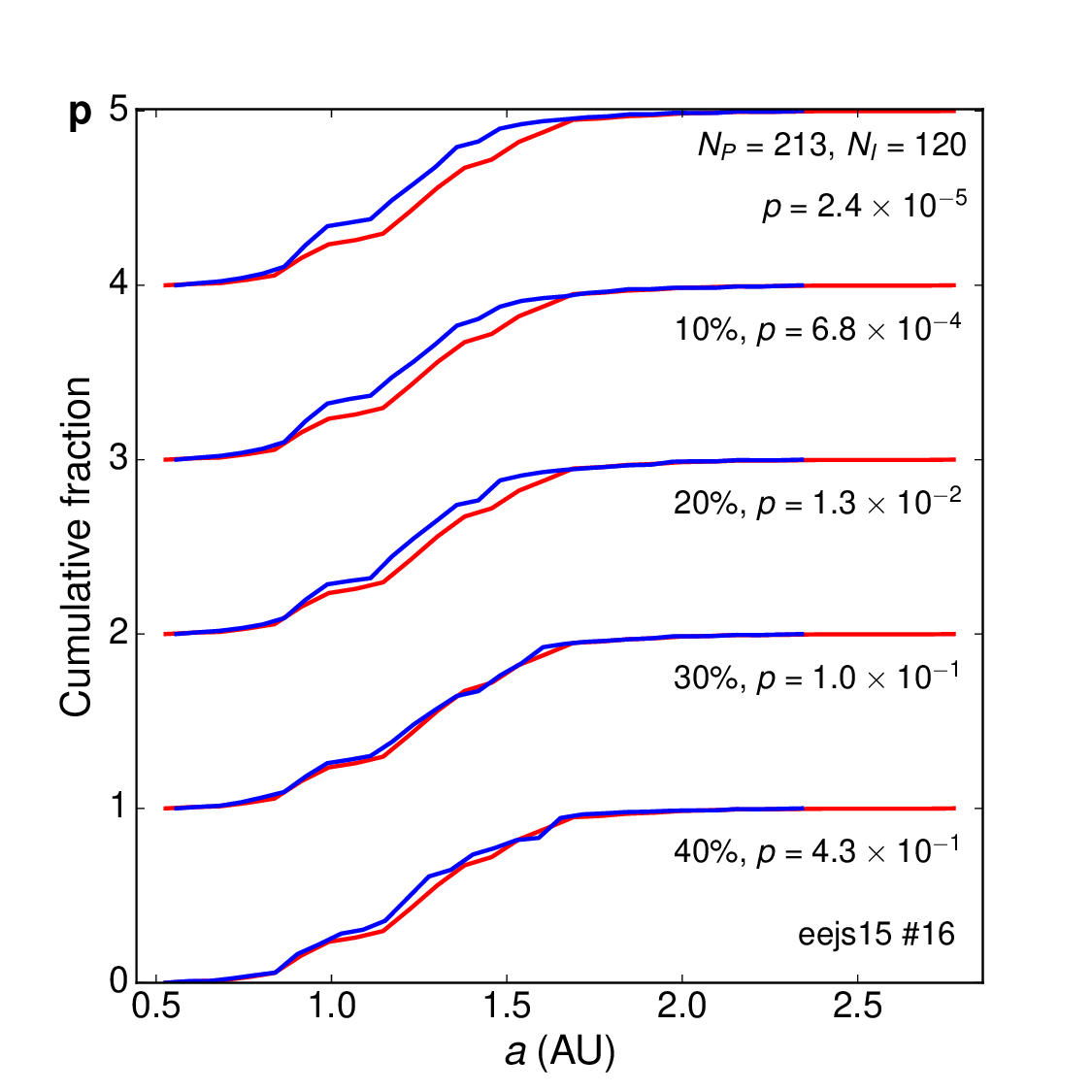}\includegraphics[scale=0.25]{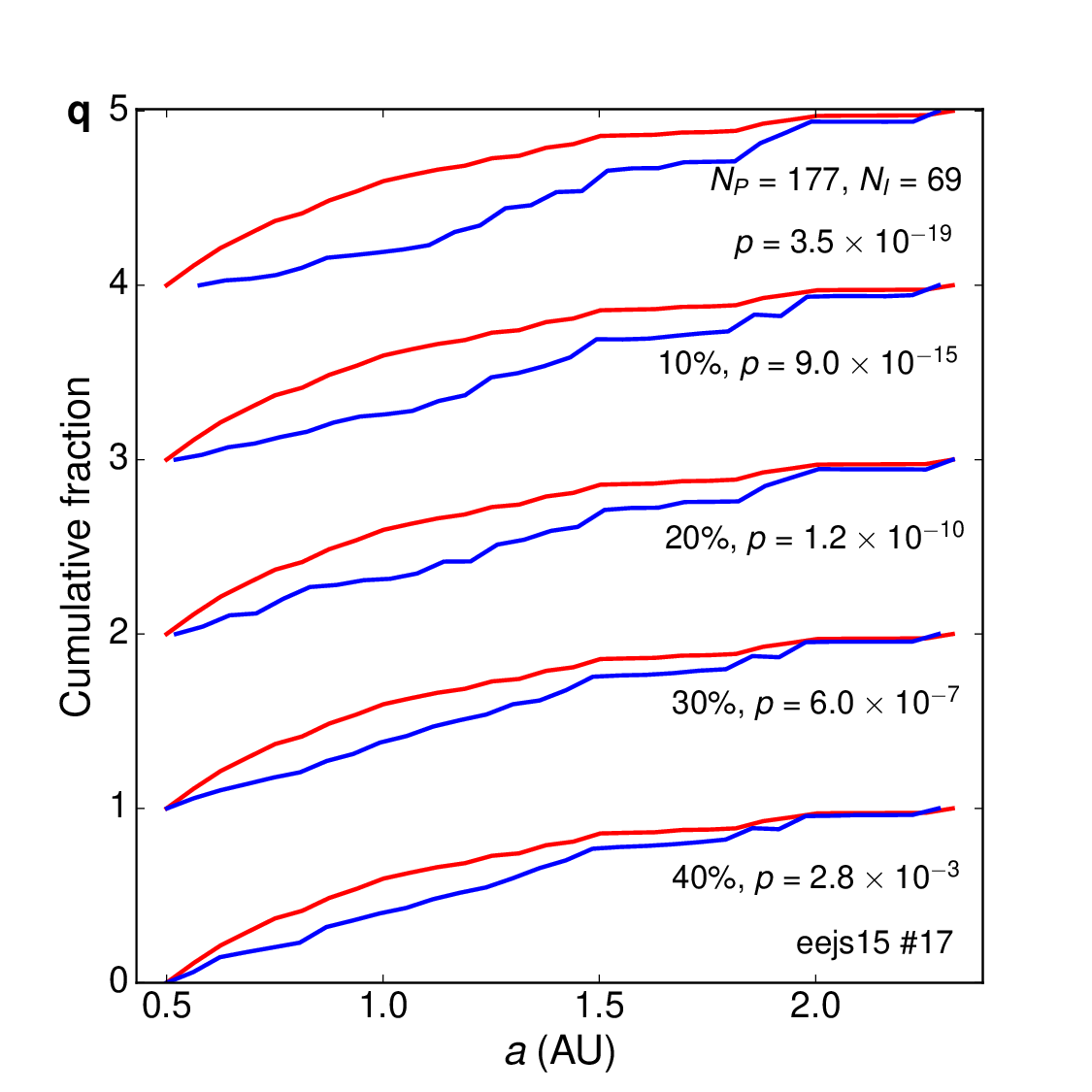}\includegraphics[scale=0.25]{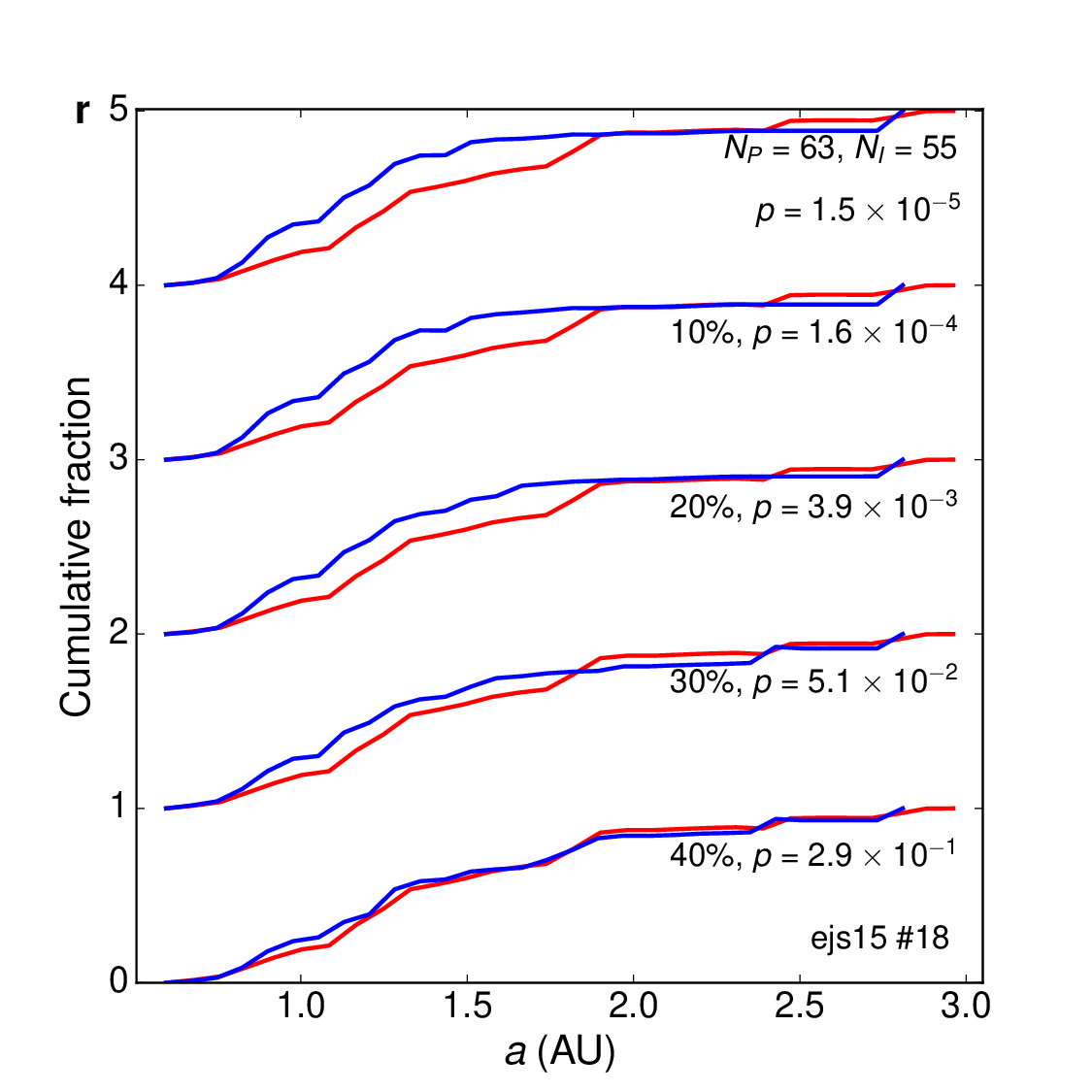}

\includegraphics[scale=0.25]{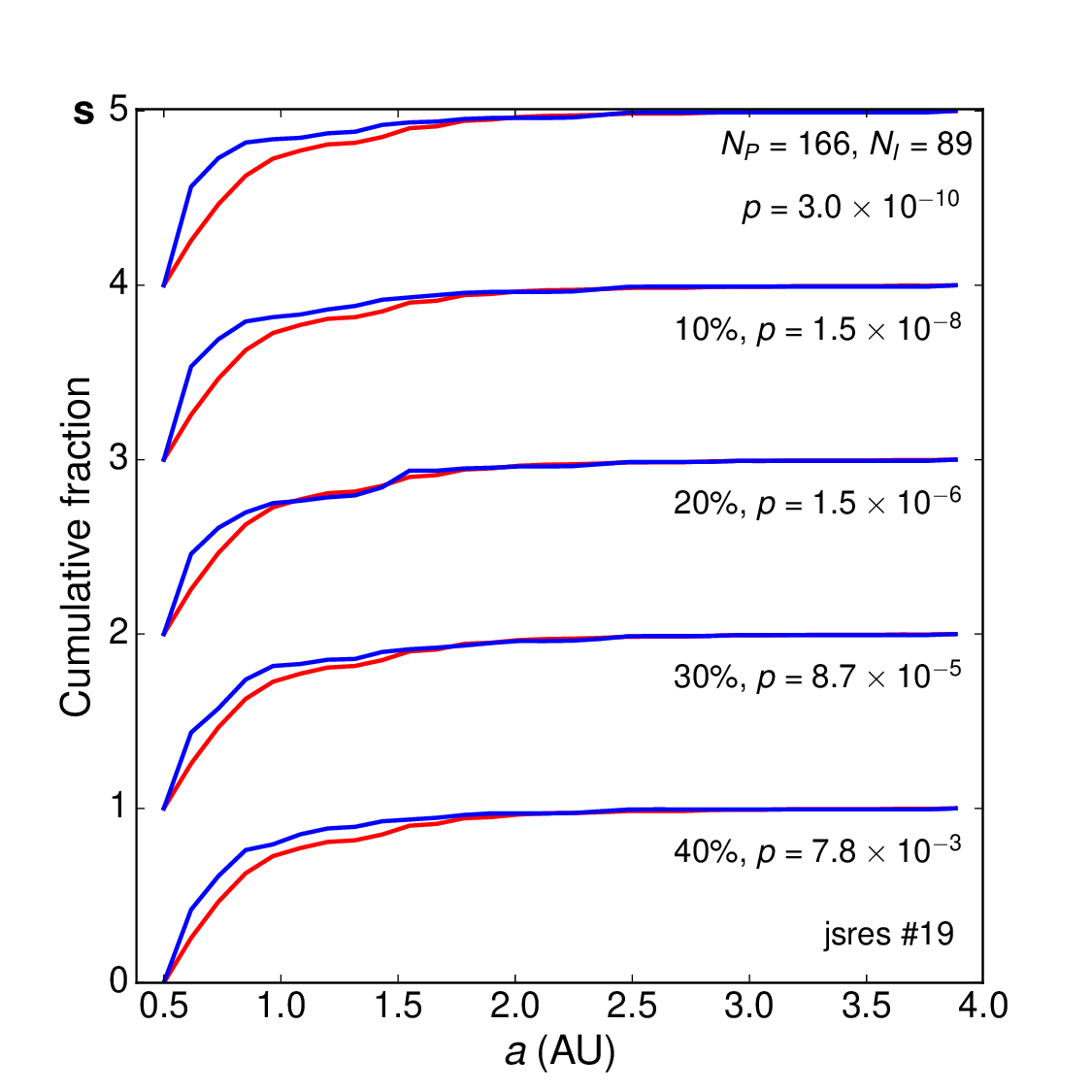}\includegraphics[scale=0.25]{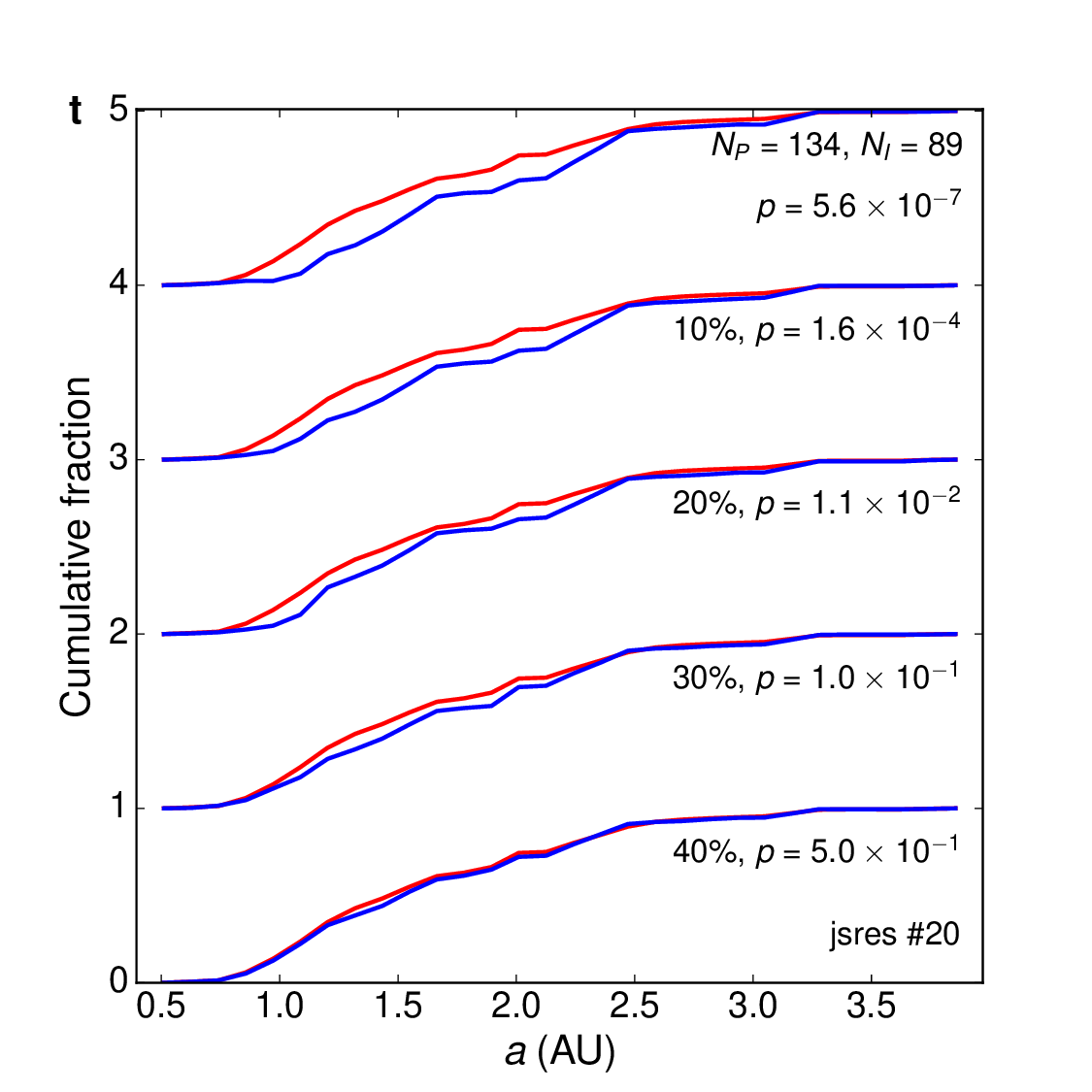}~~~~~~~~~~~~~~~~~~~~~~~~~~~~~~~~~~~~~~~~~~~~~
\caption{{\bf The cumulative distribution of the planetesimals composing the planet
and the impactor}, showing planet-impactor pairs in Table 1, cases 13-20 (panels {\bf m}-{\bf t}).}
\end{figure}

\begin{figure}
\label{fig:min_imp0}\center\center\def\figurename{Extended Data Figure}
\includegraphics[scale=0.3]{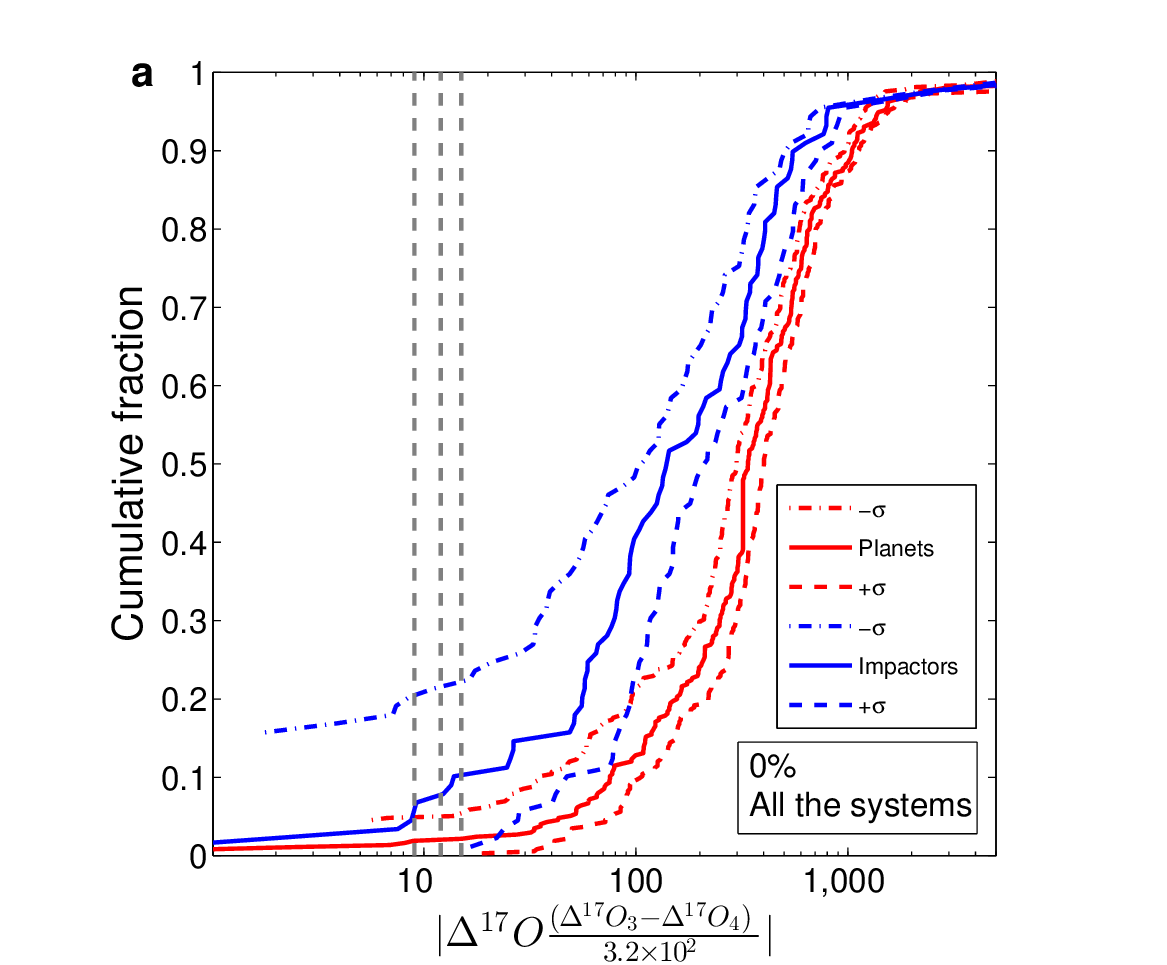}~~\includegraphics[scale=0.3]{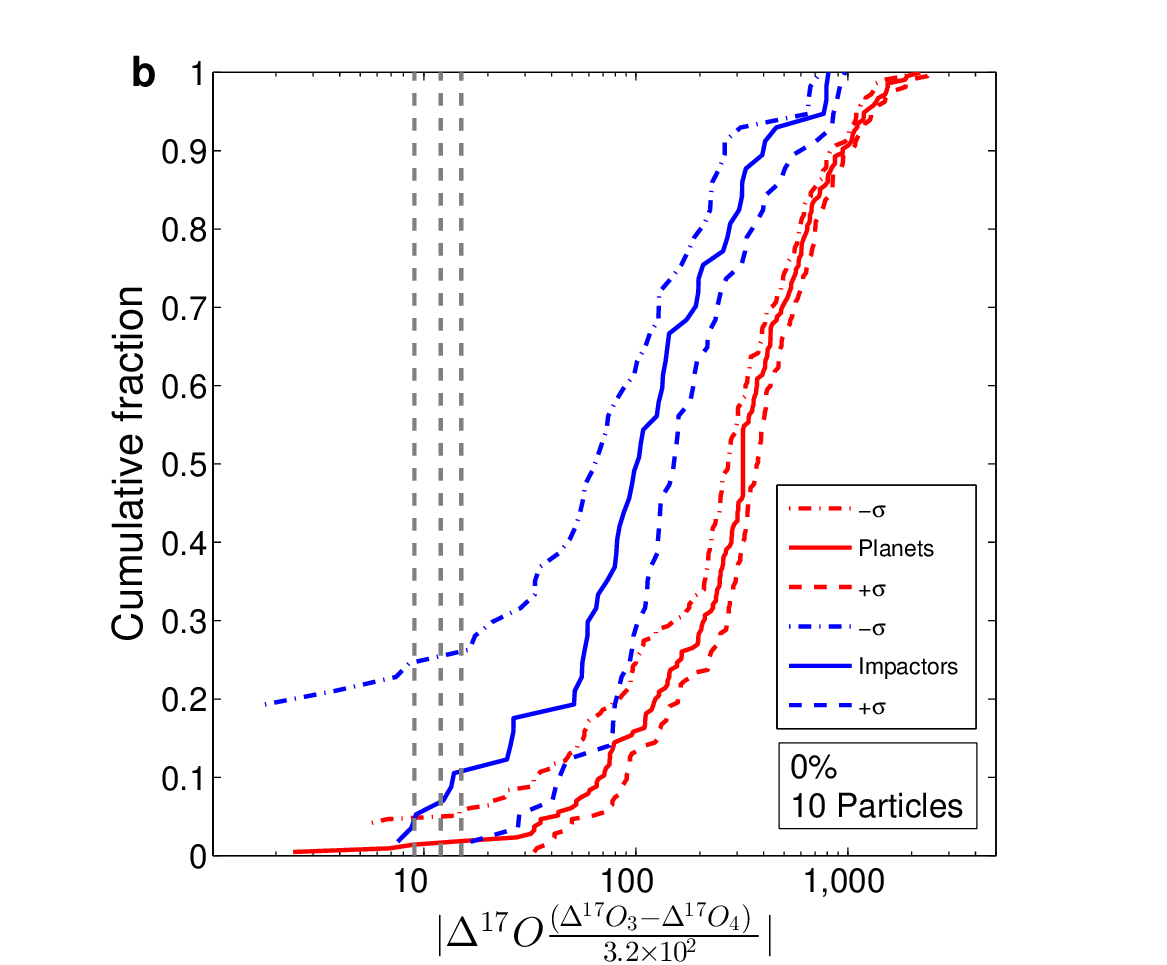}\\
\includegraphics[scale=0.3]{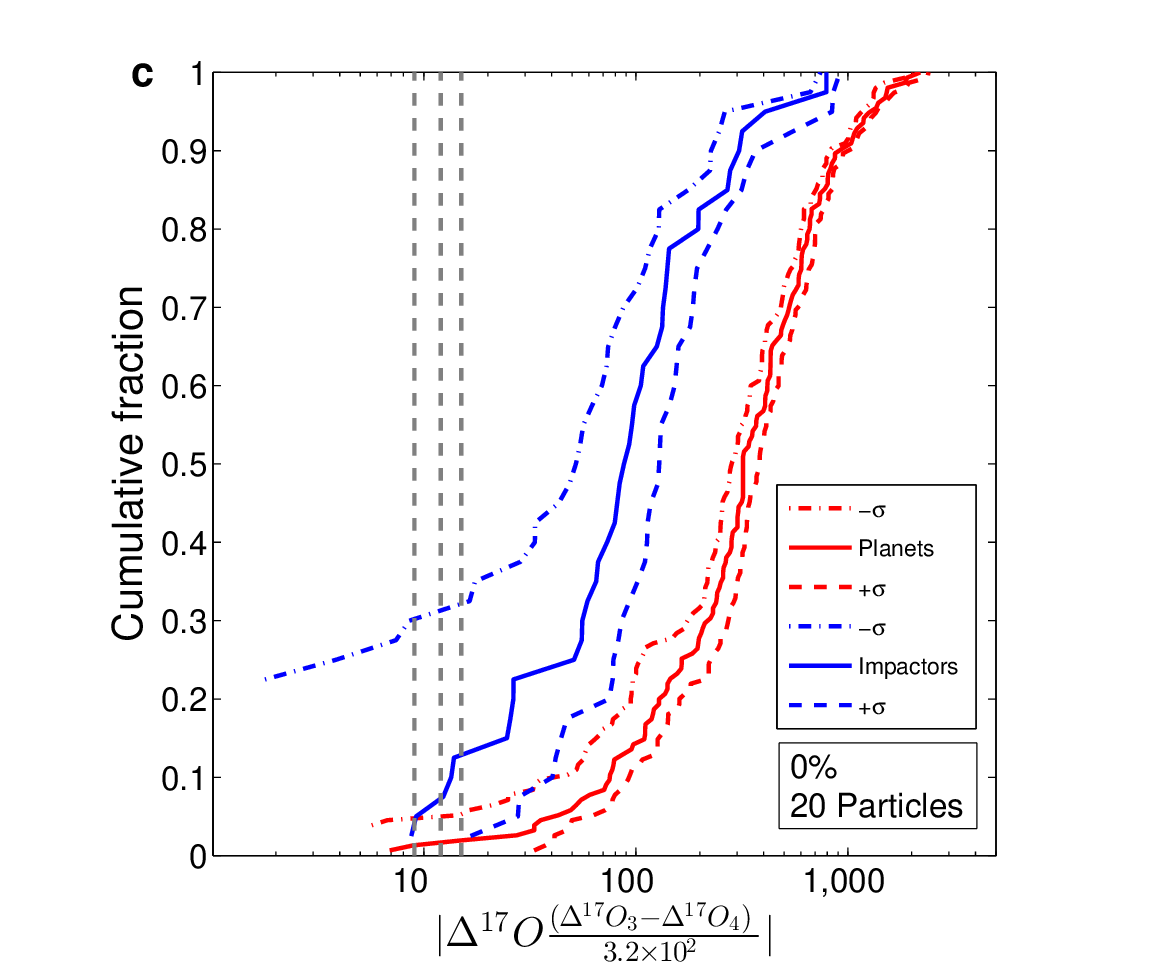}~~\includegraphics[scale=0.3]{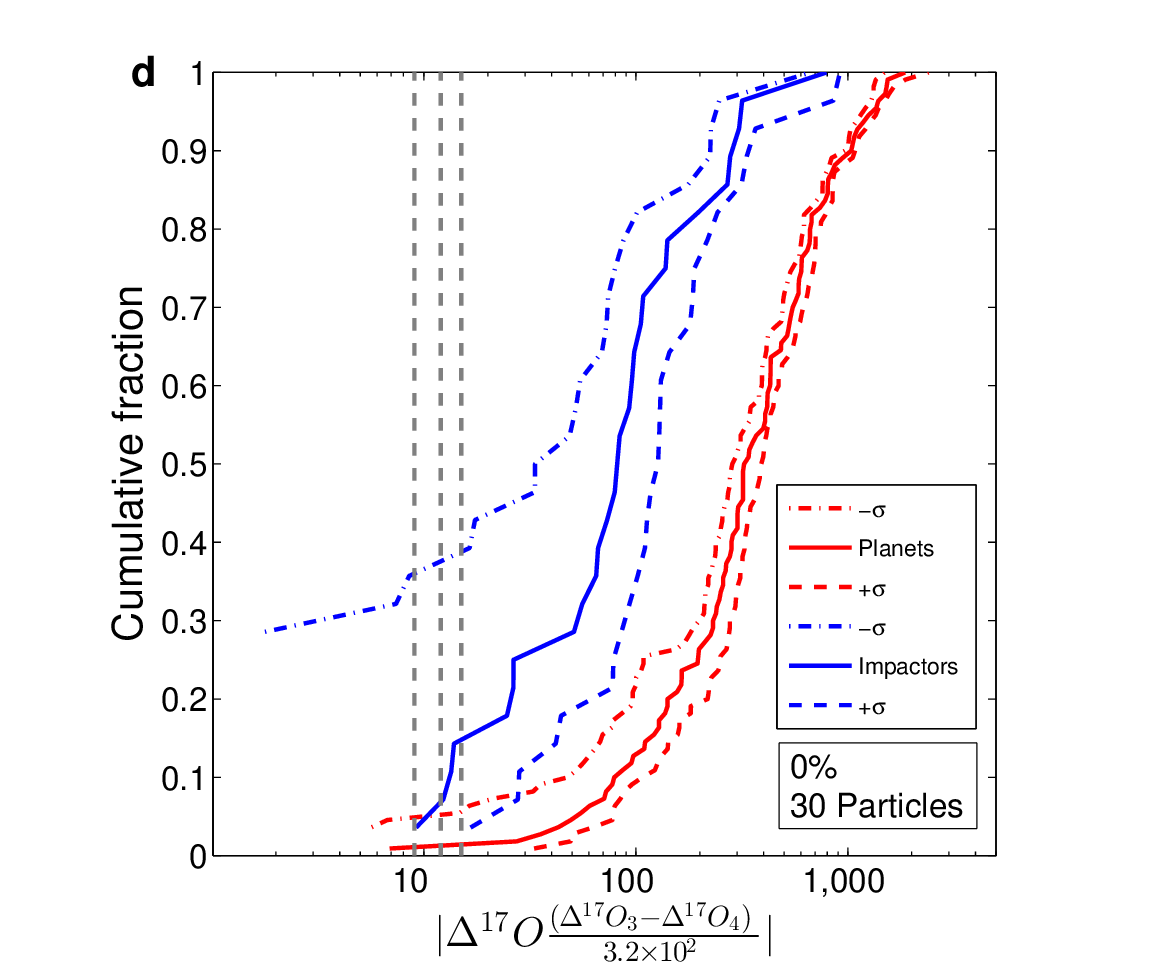}\\
\includegraphics[scale=0.3]{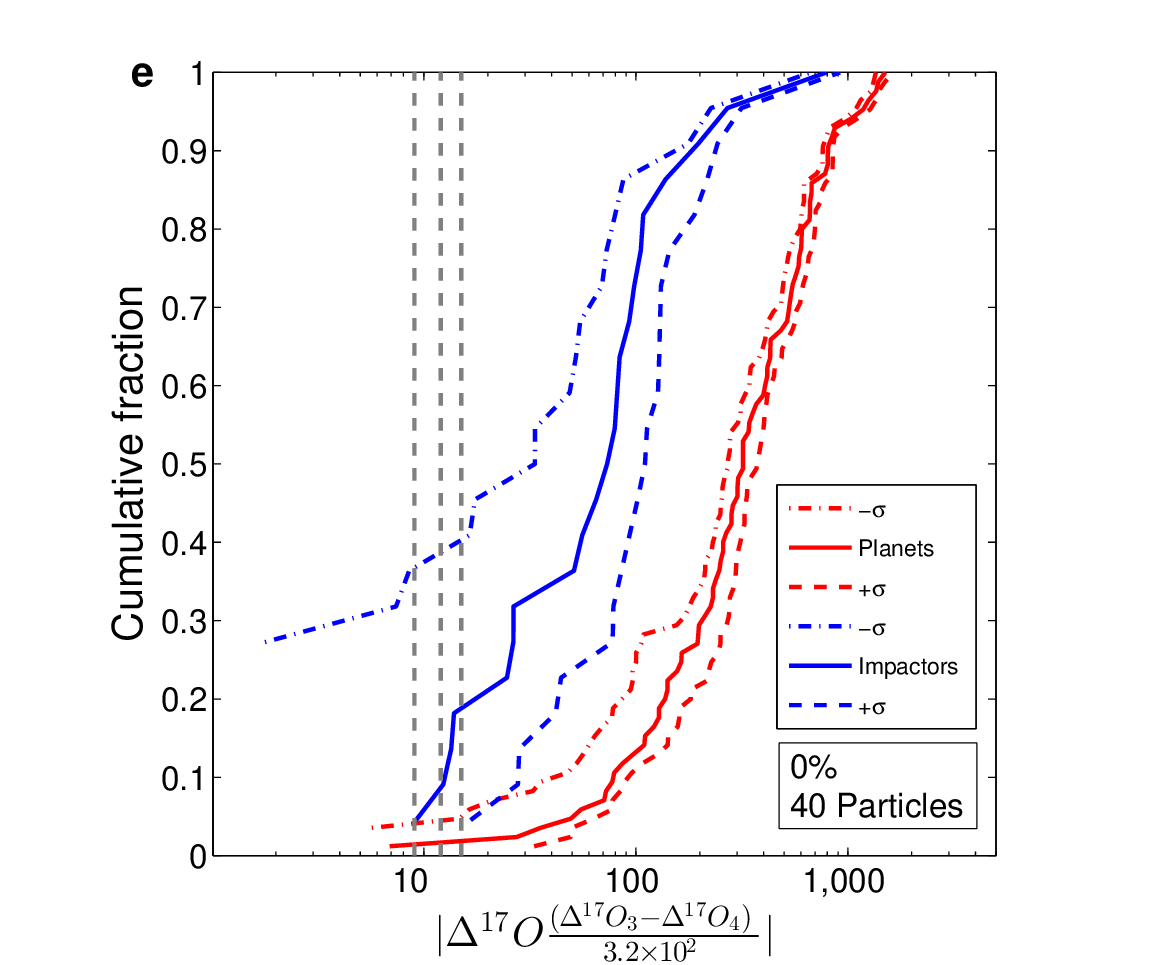}~~\includegraphics[scale=0.3]{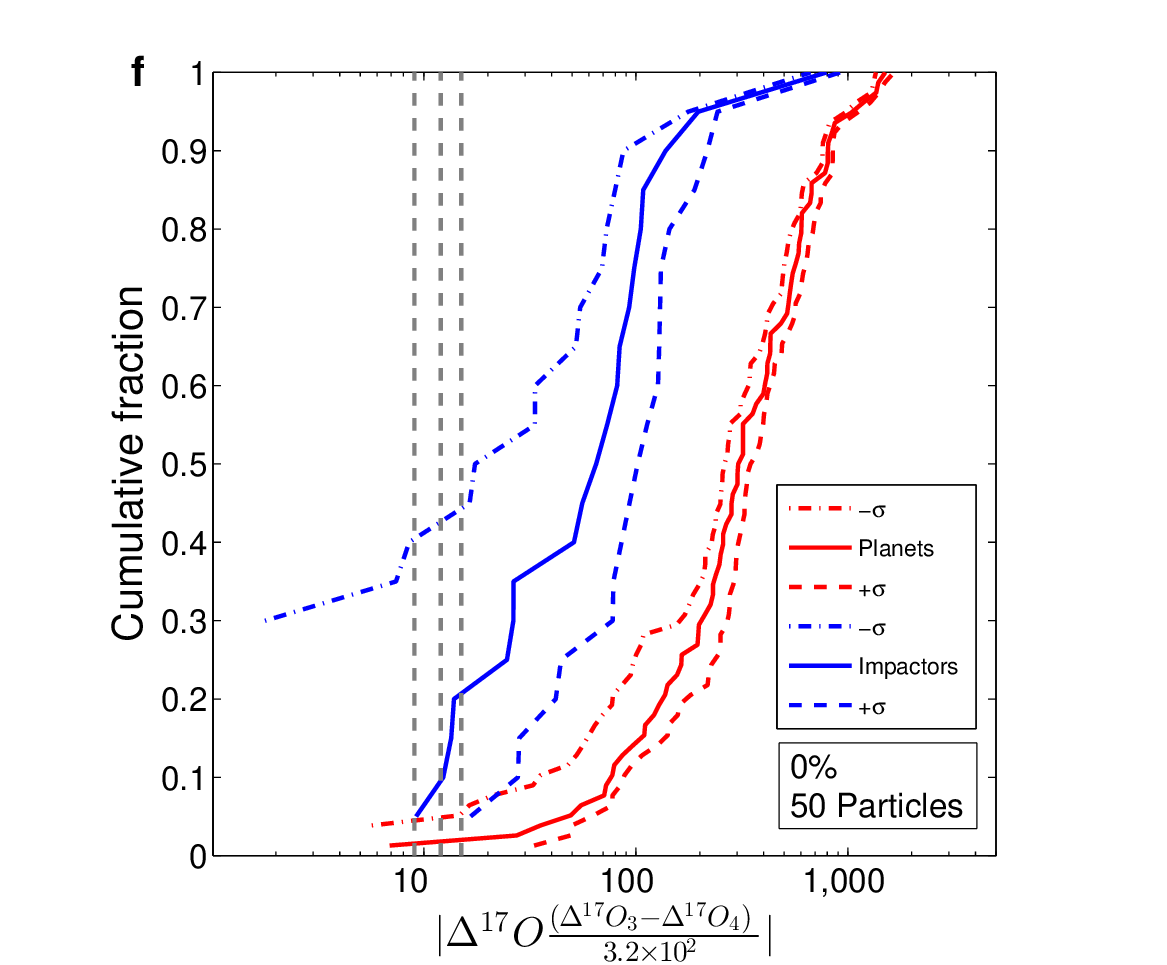}\\
\caption{{\bf The cumulative distribution of the absolute $\Delta^{17}{\rm O}$ differences between planets and their last giant impactors (blue), compared with the differences between planets in the same system (red) assuming 0\% mixing between Earth and Moon material}. From the top left panel ({\bf a}) to the bottom right panel ({\bf f})  we consider all the systems, regardless of the number of particles that contributed to their formation, and planets and last impactors composed by a minimum of 10, 20, 40 and 50 particles.  Only last-impactors with mass $>0.5$M$_{\rm Mars}$ have been taken into account.  }
\end{figure}
\begin{figure}
\label{fig:min_imp20}\center\center\def\figurename{Extended Data Figure}
\includegraphics[scale=0.3]{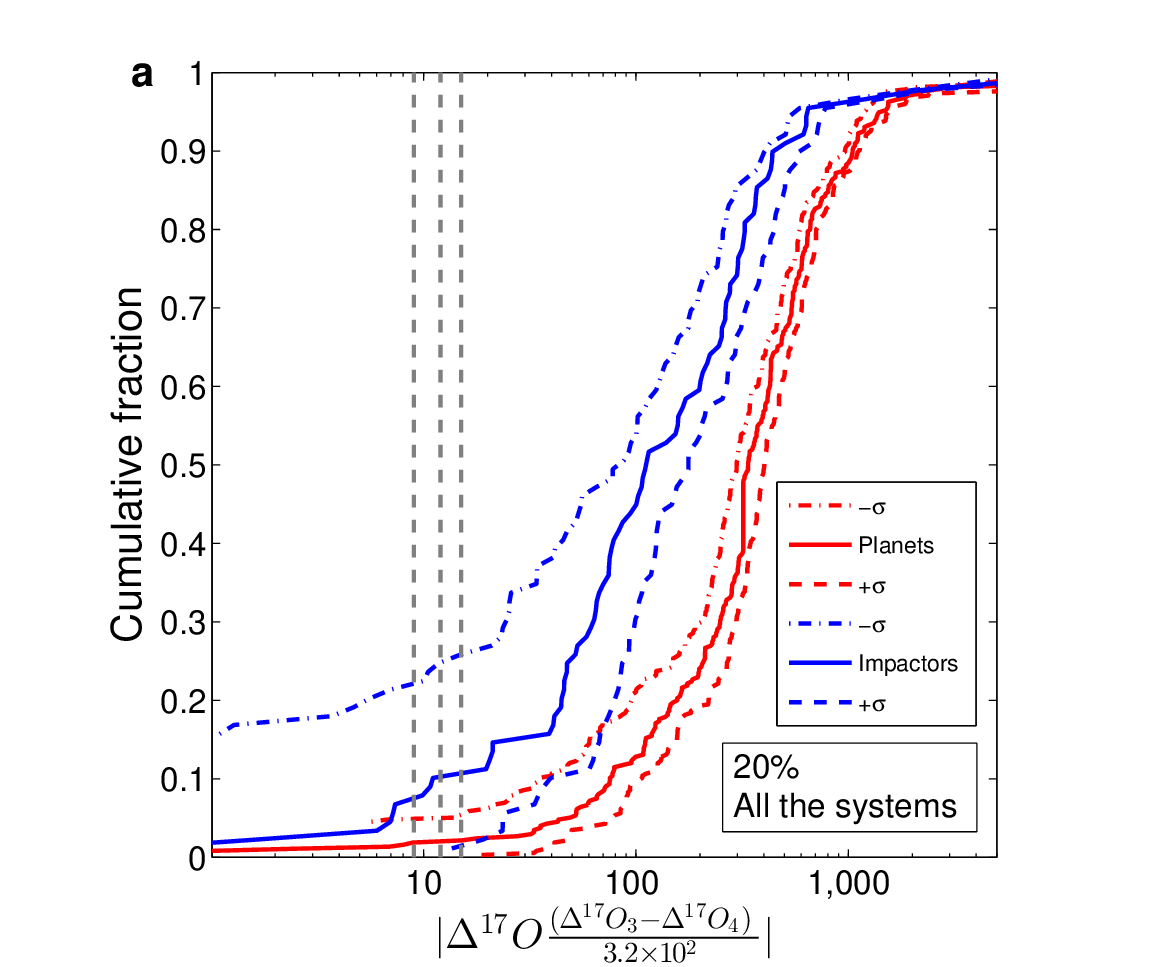}~~\includegraphics[scale=0.3]{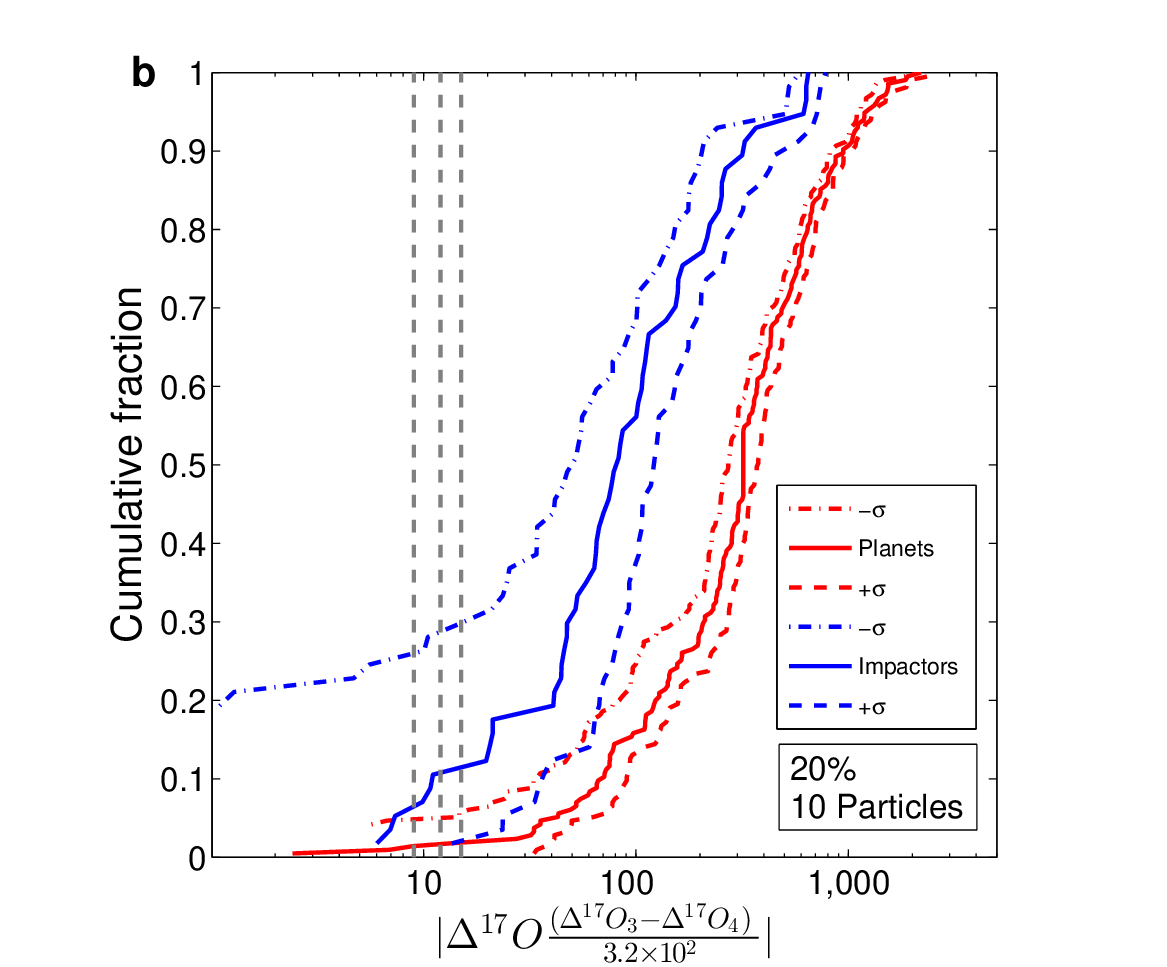}\\
\includegraphics[scale=0.3]{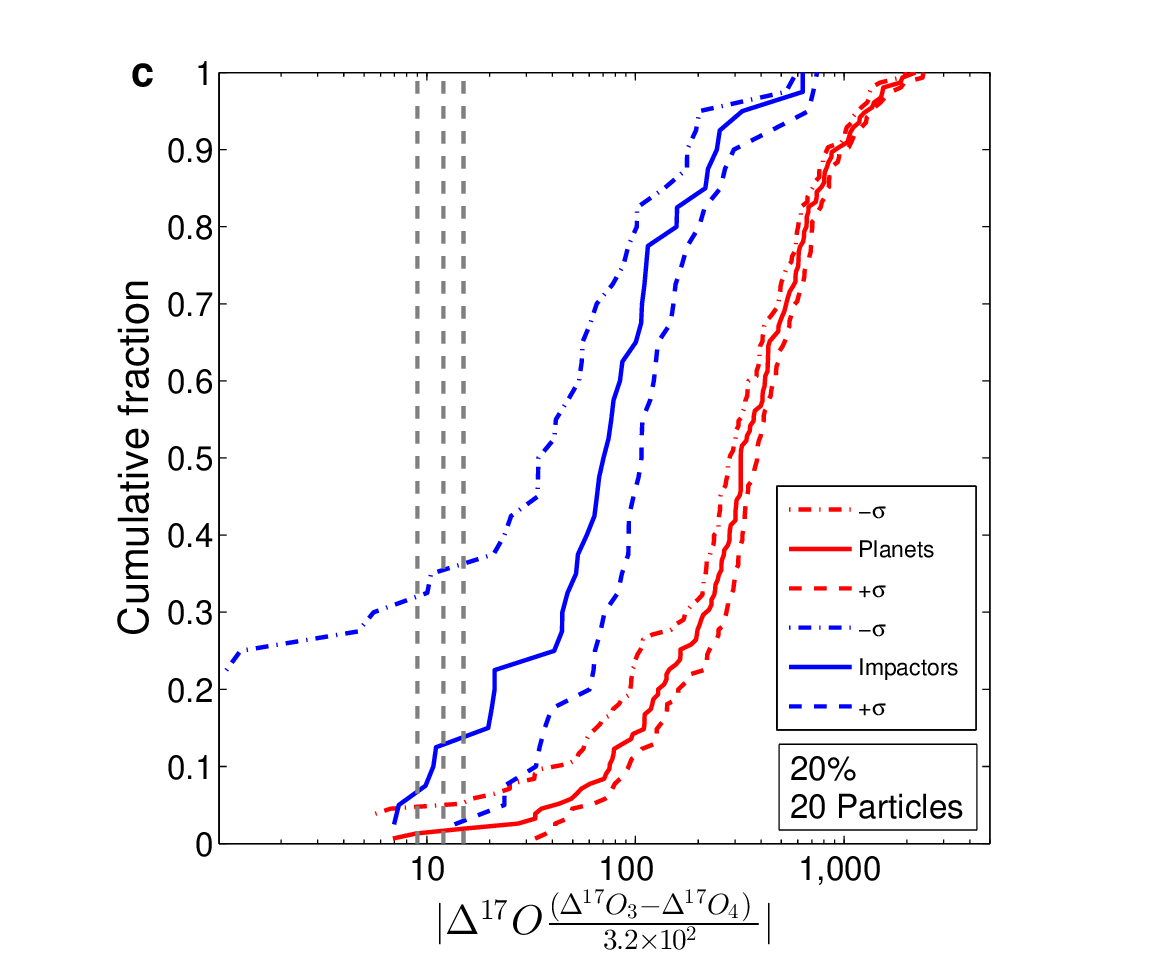}~~\includegraphics[scale=0.3]{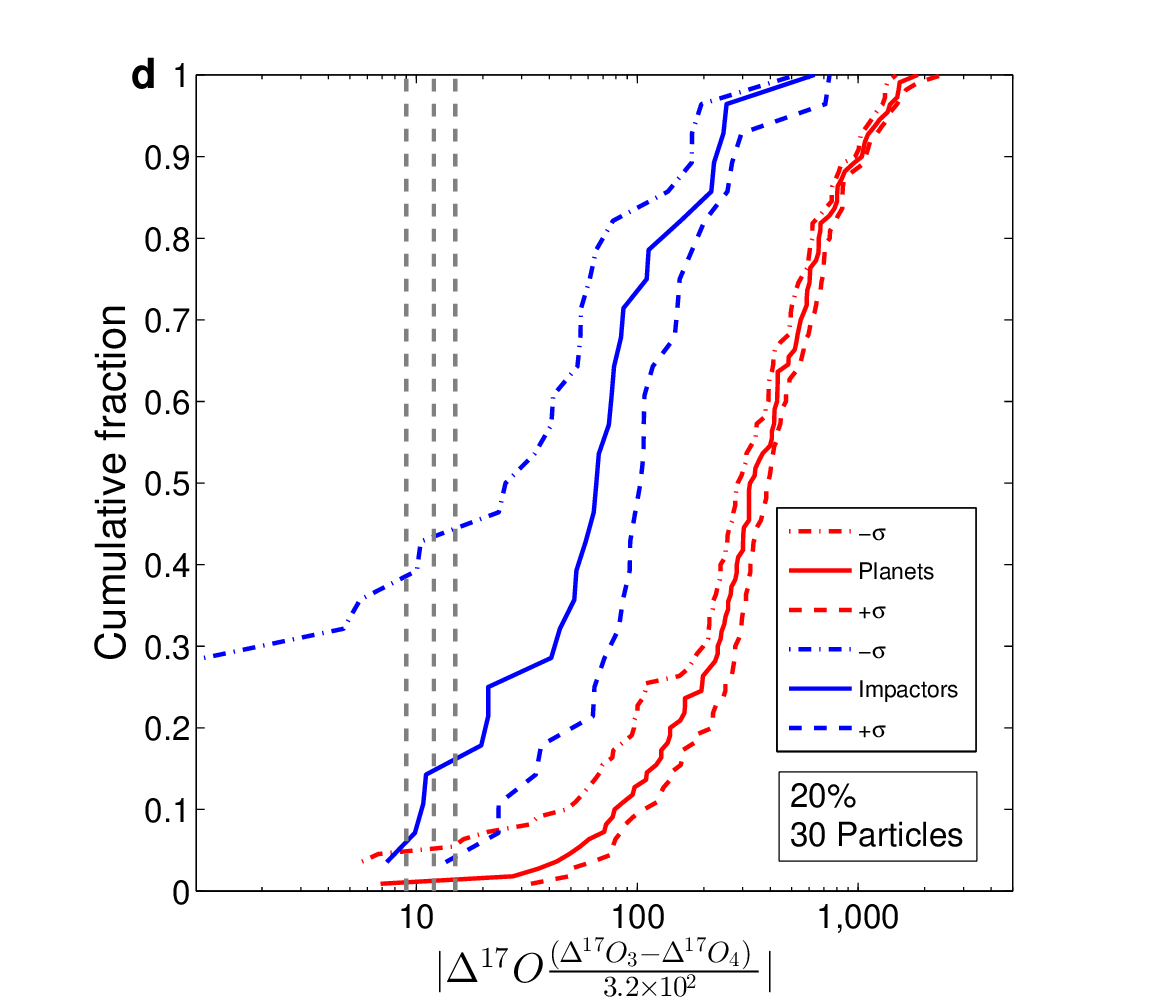}\\
\includegraphics[scale=0.3]{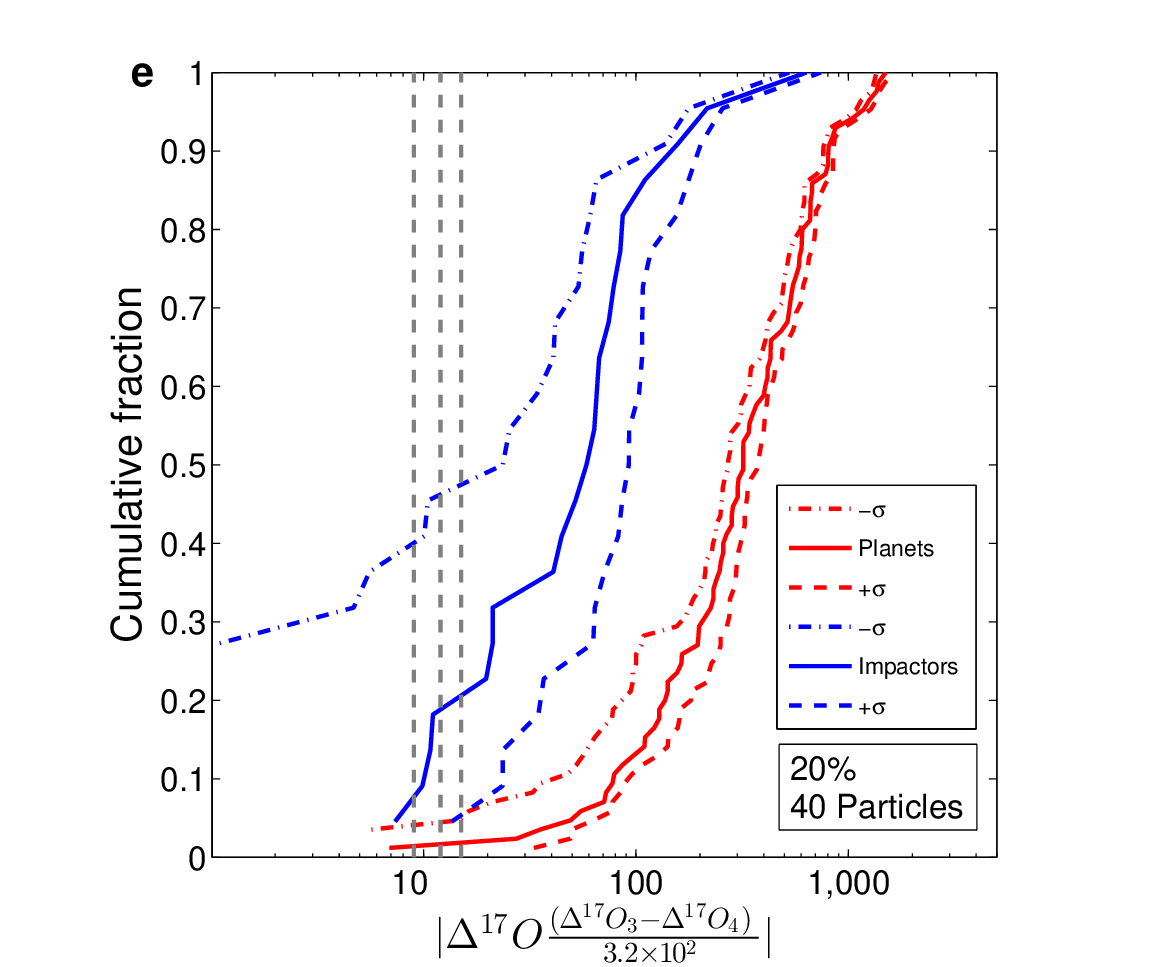}~~\includegraphics[scale=0.3]{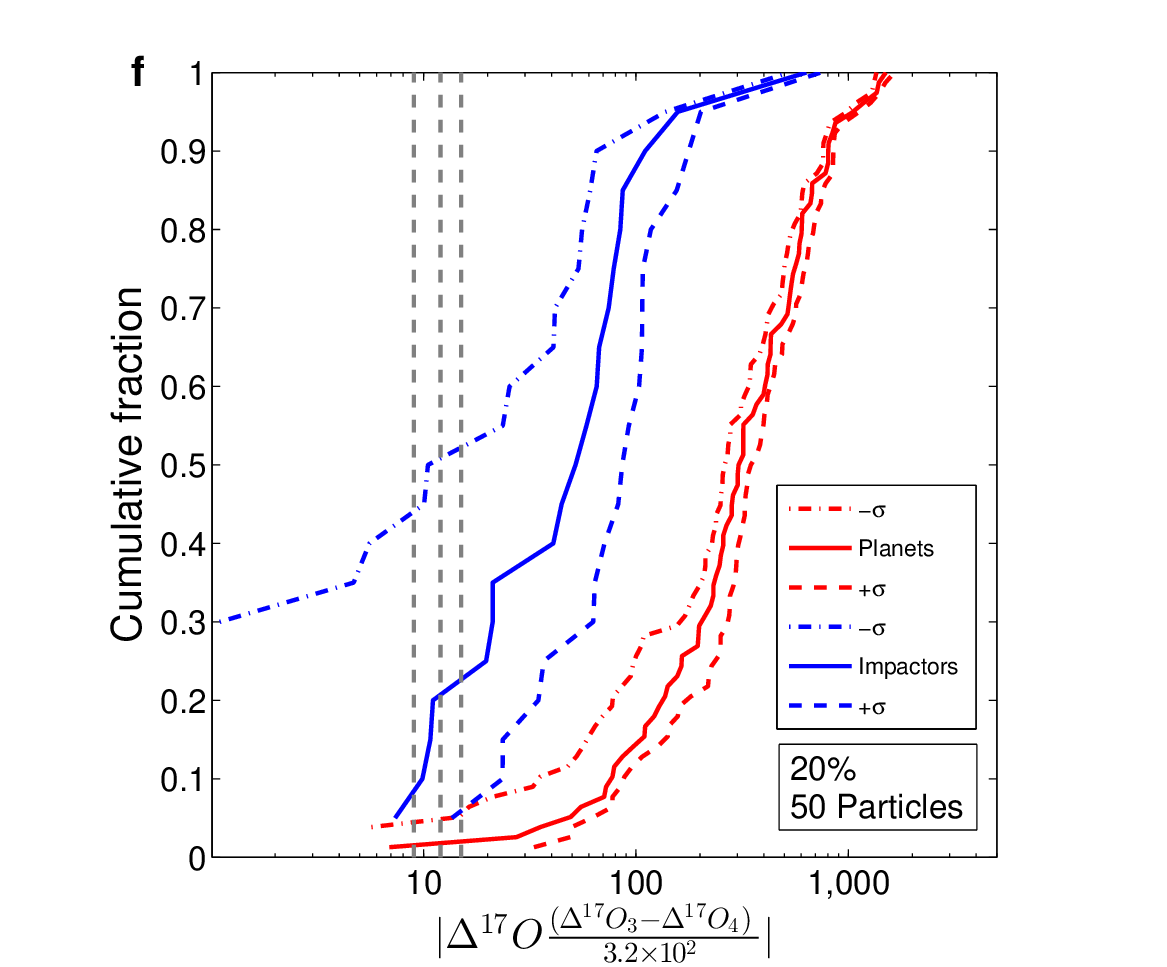}\\
\caption{{\bf The cumulative distribution of the absolute $\Delta^{17}{\rm O}$ differences between planets and their last giant impactors (blue), compared with the differences between planets in the same system (red) assuming 20\% mixing between Earth and Moon material}. From the top left panel ({\bf a})  to the bottom right panel ({\bf f})  we consider all the systems, regardless of the number of particles that contributed to their formation, and planets and last impactors composed by a minimum of 10, 20, 40 and 50 particles.   Only last-impactors with mass $>0.5$M$_{\rm Mars}$ have been taken into account.  }
\end{figure}
\begin{figure}\center\center\def\figurename{Extended Data Figure}
\label{fig:min_imp40}
\includegraphics[scale=0.3]{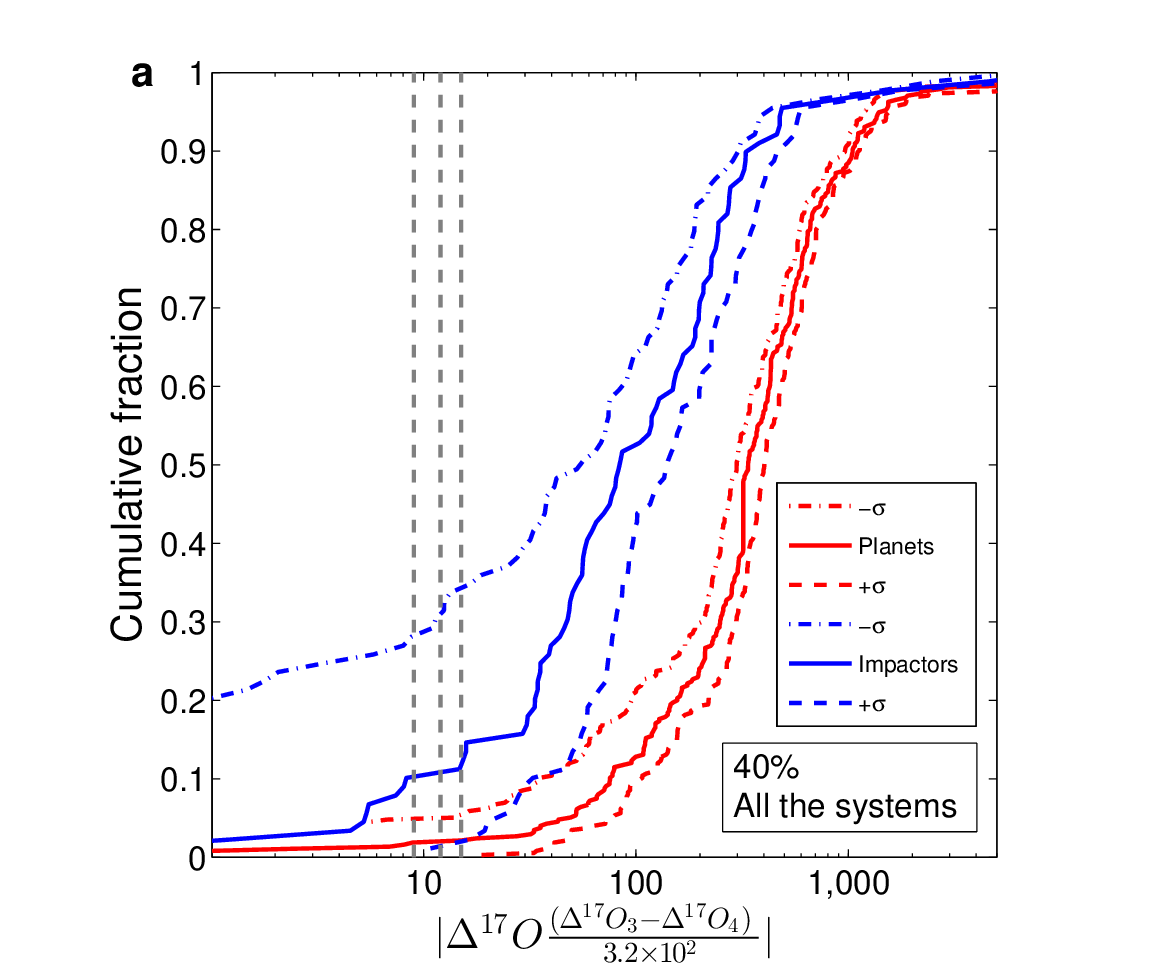}~~\includegraphics[scale=0.3]{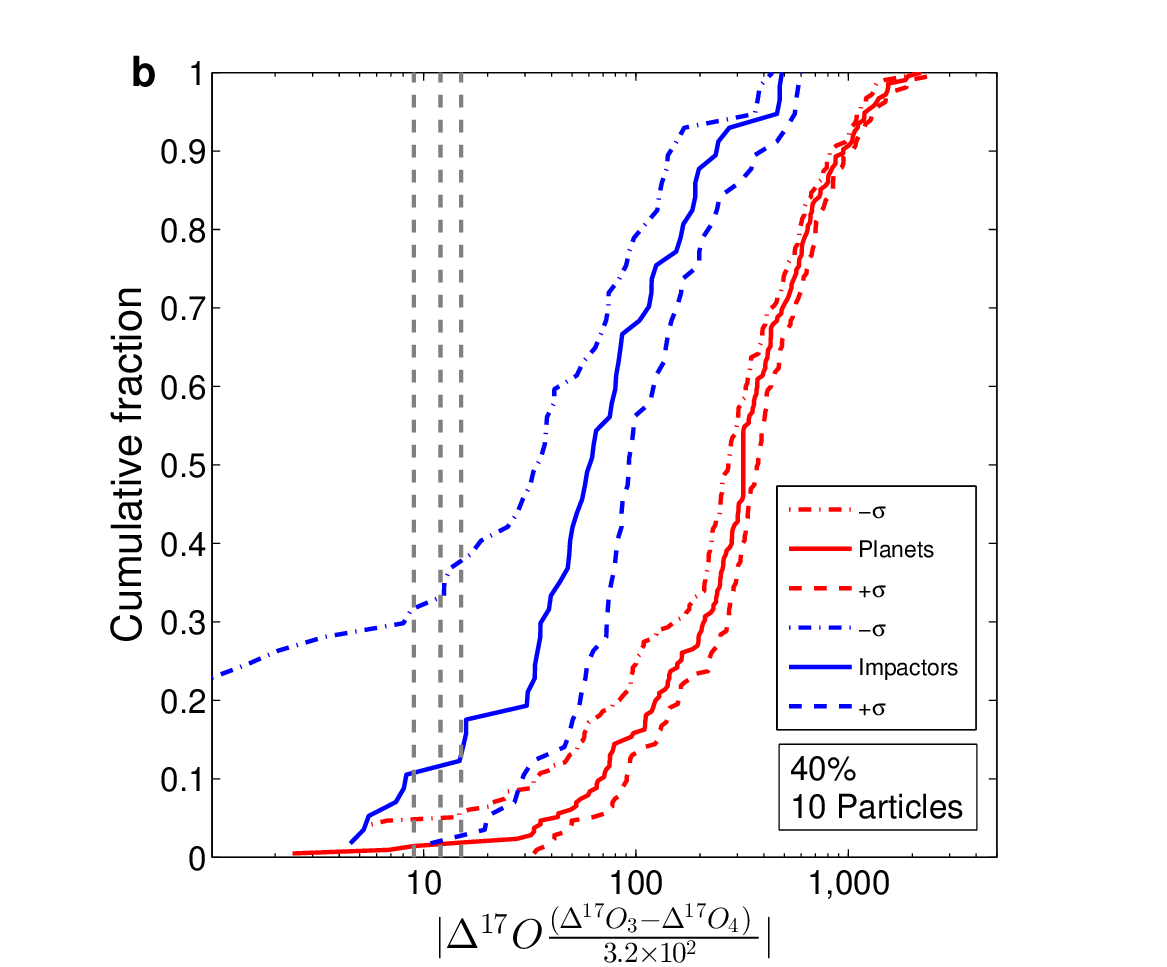}\\
\includegraphics[scale=0.3]{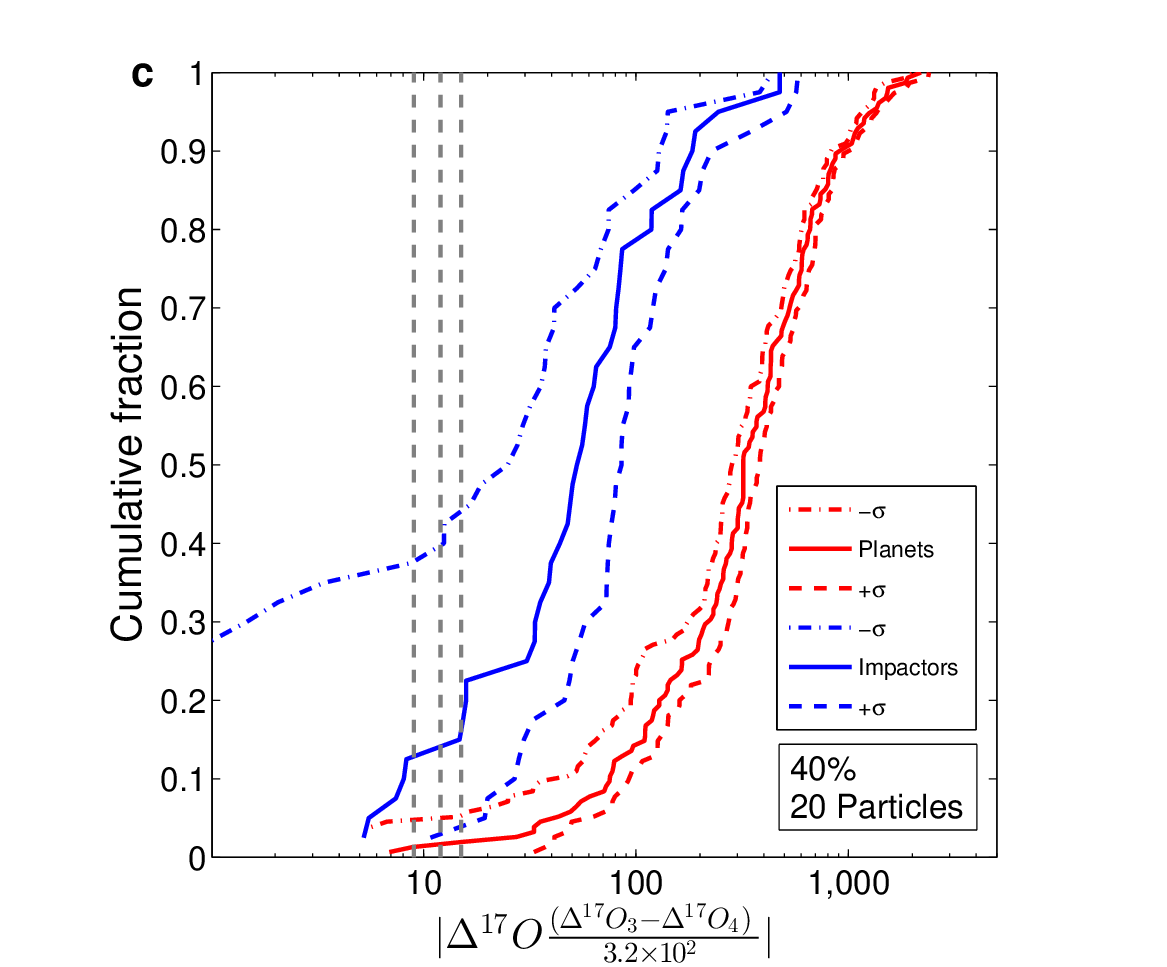}~~\includegraphics[scale=0.3]{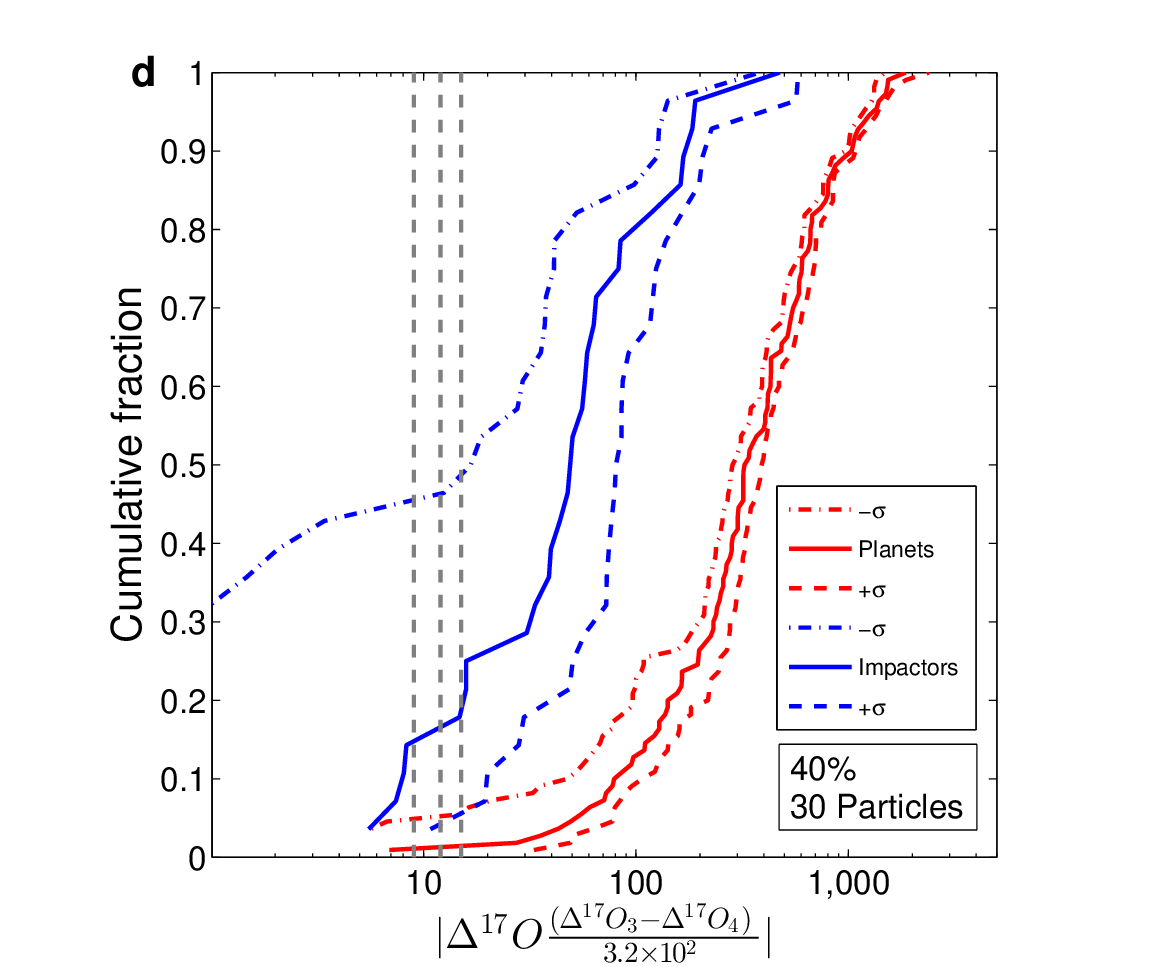}\\
\includegraphics[scale=0.3]{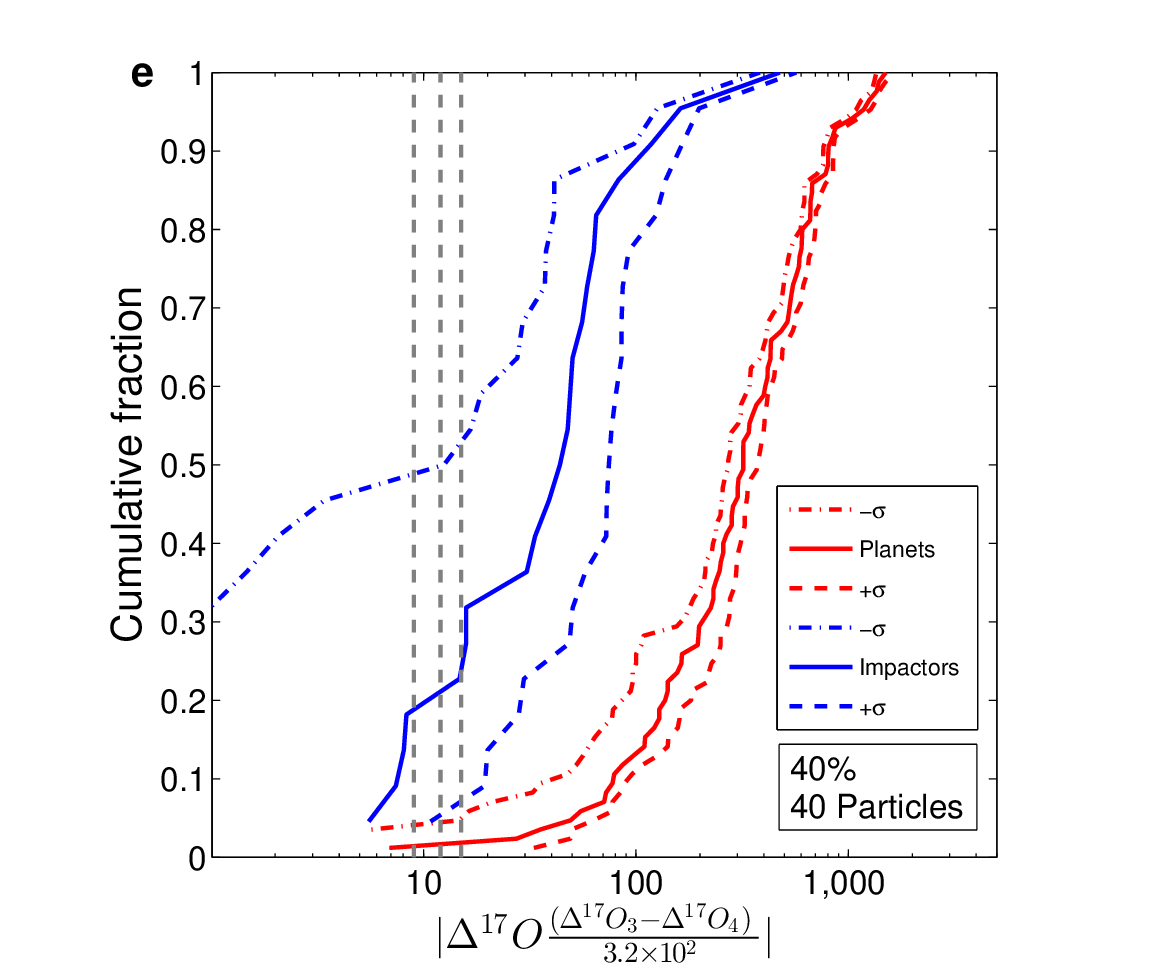}~~\includegraphics[scale=0.3]{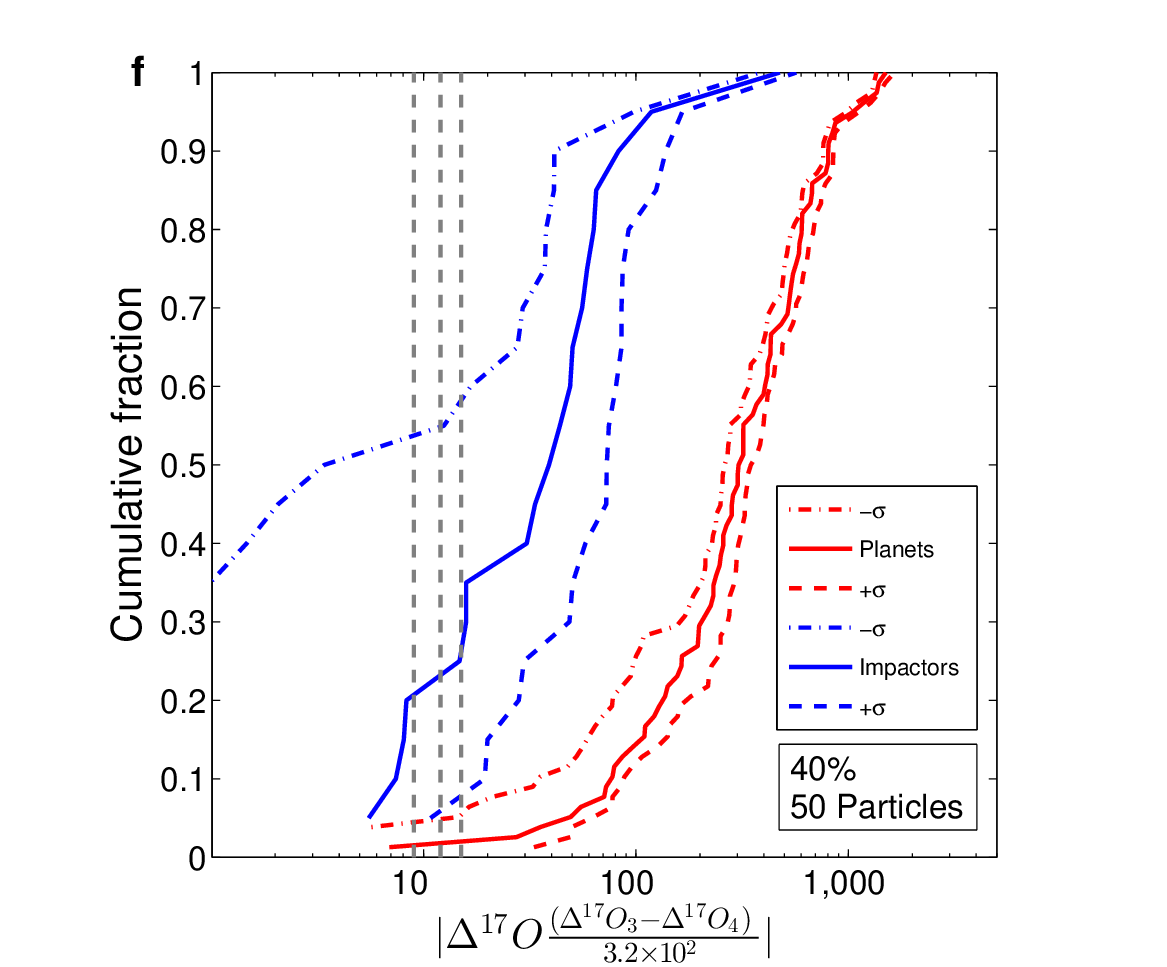}
\caption{{\bf The cumulative distribution of the absolute $\Delta^{17}{\rm O}$ differences between planets and their last giant impactors (blue), compared with the differences between planets in the same system (red) assuming 40\% mixing between Earth and Moon material}. From the top left panel ({\bf a})  to the bottom right panel ({\bf f})  we consider all the systems, regardless of the number of particles that contributed to their formation, and planets and last impactors composed by a minimum of 10, 20, 40 and 50 particles. Only last-impactors with mass $>0.5$M$_{\rm Mars}$ have been taken into account.  }
\end{figure}

\clearpage
\part*{Supplementary Information}

\begin{longtable}{|c|c|c|c|c|c|}
\hline 
\textbf{Model} & \textbf{Diff. \#} & \textbf{Mass ($M_{\oplus}$)} & $a$(AU) & \textbf{Ref. Planet} & \multicolumn{1}{c|}{\textbf{$\Delta^{17}O$ (ppm)}}\tabularnewline
\hline 
cjs15 & 1 & $3.02\times10^{-1}$ & $0.498$ & cjs15 \#1.1 & $-27\pm21$\tabularnewline
\hline 
cjs15 & 2 & $9.40\times10^{-1}$ & $0.742$ & cjs15 \#1.2 & $0\pm13$\tabularnewline
\hline 
cjs15 & 3 & $1.45$ & $1.41$ & cjs15 \#1.3 & $(-2.82\pm0.21)\times10^{2}$\tabularnewline
\hline 
cjs15 & 4 & $6.10\times10^{-2}$ & $2.06$ & cjs15 \#1.4 & $(-6.02\pm0.79)\times10^{2}$\tabularnewline
\hline 
cjs15 & 5 & $3.68\times10^{-2}$ & $2.35$ & cjs15 \#1.5 & $(-5.85\pm0.49)\times10^{2}$\tabularnewline
\hline 
\hline 
cjs15 & 6 & $7.83\times10^{-1}$ & $0.580$ & cjs15 \#2.1 & $0\pm25$\tabularnewline
\hline 
cjs15 & 7 & $5.57\times10^{-1}$ & $0.891$ & cjs15 \#2.2 & $-77\pm30$\tabularnewline
\hline 
cjs15 & 8 & $9.78\times10^{-1}$ & $1.37$ & cjs15 \#2.3 & $-287\pm31$\tabularnewline
\hline 
cjs15 & 9 & $5.76\times10^{-1}$ & $1.91$ & cjs15 \#2.4 & $-606\pm46$\tabularnewline
\hline 
\hline 
cjs1 & 10 & $1.25$ & $0.720$ & cjs1 \#3.1 & $0\pm7.2$\tabularnewline
\hline 
cjs1 & 11 & $1.05$ & $1.94$ & cjs1 \#3.2 & $-231\pm19$\tabularnewline
\hline 
cjs1 & 12 & $8.80\times10^{-2}$ & $2.33$ & cjs1 \#3.3 & $-109\pm31$\tabularnewline
\hline 
cjs1 & 13 & $2.47\times10^{-2}$ & $2.45$ & cjs1 \#3.4 & $-429\pm23$\tabularnewline
\hline 
\hline 
cjs1 & 14 & $1.25$ & $0.720$ & cjs1 \#4.1 & $231\pm19$\tabularnewline
\hline 
cjs1 & 15 & $1.05$ & $1.94$ & cjs1 \#4.2 & $0\pm25$\tabularnewline
\hline 
cjs1 & 16 & $8.80\times10^{-2}$ & $2.33$ & cjs1 \#4.3 & $122\pm35$\tabularnewline
\hline 
cjs1 & 17 & $2.47\times10^{-2}$ & $2.45$ & cjs1 \#4.4 & $-198\pm28$\tabularnewline
\hline 
\hline 
cjs1 & 18 & $1.21$ & $0.765$ & cjs1 \#5.1 & $0\pm14$\tabularnewline
\hline 
cjs1 & 19 & $1.32$ & $1.83$ & cjs1 \#5.2 & $406\pm28$\tabularnewline
\hline 
cjs1 & 20 & $2.25\times10^{-2}$ & $3.55$ & cjs1 \#5.3 & $86.4\pm9.3$\tabularnewline
\hline 
\hline 
cjsecc & 21 & $6.85\times10^{-1}$ & $0.574$ & cjsecc \#6.1 & $128\pm32$\tabularnewline
\hline 
cjsecc & 22 & $9.35\times10^{-1}$ & $0.901$ & cjsecc \#6.2 & $0\pm28$\tabularnewline
\hline 
cjsecc & 23 & $5.83\times10^{-1}$ & $1.41$ & cjsecc \#6.3 & $-211\pm33$\tabularnewline
\hline 
cjsecc & 24 & $1.61\times10^{-1}$ & $2.45$ & cjsecc \#6.4 & $-531\pm36$\tabularnewline
\hline 
\hline 
cjsecc  & 25 & $1.01$ & $0.557$ & cjsecc \#7.1 & $0\pm12$\tabularnewline
\hline 
cjsecc  & 26 & $7.18\times10^{-1}$ & $0.953$ & cjsecc \#7.2 & $72\pm12$\tabularnewline
\hline 
cjsecc & 27 & $5.93\times10^{-1}$ & $1.52$ & cjsecc \#7.3 & $-49\pm14$\tabularnewline
\hline 
cjsecc & 28 & $4.99\times10^{-2}$ & $1.98$ & cjsecc \#7.4 & $(-3.69\pm0.43)\times10^{2}$\tabularnewline
\hline 
cjsecc & 29 & $4.86\times10^{-2}$ & $2.36$ & cjsecc \#7.5 & $96\pm39$\tabularnewline
\hline 
cjsecc & 30 & $3.21\times10^{-2}$ & $2.51$ & cjsecc \#7.6 & $-320.9\pm8.3$\tabularnewline
\hline 
cjsecc & 31 & $2.21\times10^{-1}$ & $2.76$ & cjsecc \#7.7 & $-283\pm52$\tabularnewline
\hline 
\hline 
cjsecc  & 32 & $1.02$ & $0.622$ & cjsecc \#8.1 & $0\pm38$\tabularnewline
\hline 
cjsecc  & 33 & $5.61\times10^{-1}$ & $1.13$ & cjsecc \#8.2 & $-228\pm46$\tabularnewline
\hline 
cjsecc  & 34 & $6.92\times10^{-1}$ & $1.38$ & cjsecc \#8.3 & $-585\pm47$\tabularnewline
\hline 
cjsecc  & 35 & $1.40\times10^{-1}$ & $2.12$ & cjsecc \#8.4 & $-905\pm52$\tabularnewline
\hline 
cjsecc  & 36 & $3.59\times10^{-2}$ & $2.81$ & cjsecc \#8.5 & $(1.0\pm1.2)\times10^{2}$\tabularnewline
\hline 
\hline 
eejs15 & 37 & $6.97\times10^{-1}$ & $0.565$ & eejs15 \#9.1 & $0\pm4.5$\tabularnewline
\hline 
eejs15  & 38 & $8.15\times10^{-1}$ & $0.905$ & eejs15 \#9.2 & $-70.8\pm6.7$\tabularnewline
\hline 
eejs15  & 39 & $1.58\times10^{-1}$ & $1.61$ & eejs15 \#9.3 & $-79.9\pm9.4$\tabularnewline
\hline 
eejs15 & 40 & $5.20\times10^{-2}$ & $3.15$ & eejs15 \#9.4 & $-398.9\pm3.2$\tabularnewline
\hline 
\hline 
eejs15 & 41 & $5.47\times10^{-1}$ & $0.540$ & eejs15 \#10.1 & $0\pm20$\tabularnewline
\hline 
eejs15 & 42 & $7.77\times10^{-1}$ & $0.852$ & eejs15 \#10.2 & $(3.01\pm0.24)\times10^{2}$\tabularnewline
\hline 
eejs15 & 43 & $1.06\times10^{-1}$ & $1.20$ & eejs15 \#10.3 & $(8.69\pm0.24)\times10^{2}$\tabularnewline
\hline 
eejs15 & 44 & $3.44\times10^{-1}$ & $1.51$ & eejs15 \#10.4 & $(5.48\pm0.47)\times10^{2}$\tabularnewline
\hline 
eejs15 & 45 & $5.00\times10^{-2}$ & $1.61$ & eejs15 \#10.5 & $(8.41\pm0.23)\times10^{2}$\tabularnewline
\hline 
\hline 
eejs15 & 46 & $5.47\times10^{-1}$ & $0.540$ & eejs15 \#11.1 & $(-3.01\pm0.24)\times10^{2}$\tabularnewline
\hline 
eejs15 & 47 & $7.77\times10^{-1}$ & $0.852$ & eejs15 \#11.2 & $0\pm27$\tabularnewline
\hline 
eejs15 & 48 &$1.06\times10^{-1}$ & $1.20$   & eejs15 \#11.3 & $(5.67\pm0.78)\times10^{2}$\tabularnewline
\hline 
eejs15 & 49 & $3.44\times10^{-1}$ & $1.51$  & eejs15 \#11.4 & $(2.47\pm0.49)\times10^{2}$\tabularnewline
\hline 
eejs15 & 50 &$5.00\times10^{-2}$ & $1.61$  & eejs15 \#11.5 & $(5.40\pm0.26)\times10^{2}$\tabularnewline
\hline 
\hline 
eejs15 & 51 & $7.34\times10^{-1}$ & $0.590$ & eejs15 \#12.1 & $0\pm14$\tabularnewline
\hline 
eejs15 & 52 & $6.77\times10^{-1}$ & $0.917$ & eejs15 \#12.2 & $-35\pm14$\tabularnewline
\hline 
eejs15 & 53 & $4.39\times10^{-1}$ & $1.64$ & eejs15 \#12.3 & $-110\pm16$\tabularnewline
\hline 
eejs15 & 54 & $5.92\times10^{-2}$ & $3.10$ & eejs15 \#12.4 & $-430\pm17$\tabularnewline
\hline 
eejs15 &55 & $6.06\times10^{-2}$ & $3.92$ & eejs15 \#12.5 & $-604.8\pm9.7$\tabularnewline
\hline 
\hline 
eejs15 & 56 & $1.30$ & $0.654$ & eejs15 \#13.1 & $0\pm99$\tabularnewline
\hline 
eejs15 & 57 & $2.97\times10^{-1}$ & $1.37$ & eejs15 \#13.2 & $(1.6\pm1.4)\times10^{2}$\tabularnewline
\hline 
eejs15 & 58 & $3.16\times10^{-1}$ & $1.75$ & eejs15 \#13.3 & $(-1.6\pm1.5)\times10^{2}$\tabularnewline
\hline 
\hline 
eejs15 & 59 & $3.34\times10^{-1}$ & $0.518$ & eejs15 \#14.1 & $(2.25\pm0.70)\times10^{2}$\tabularnewline
\hline 
eejs15 & 60 & $5.68\times10^{-1}$ & $0.692$ & eejs15 \#14.2 & $164\pm56$\tabularnewline
\hline 
eejs15 & 61 & $4.95\times10^{-1}$ & $0.925$ & eejs15 \#14.3 & $0\pm54$\tabularnewline
\hline 
eejs15 & 62 & $2.11\times10^{-1}$ & $1.26$ & eejs15 \#14.4 & $(-3.2\pm0.70)\times10^{2}$\tabularnewline
\hline 
eejs15 & 63 & $2.26\times10^{-1}$ & $1.59$ & eejs15 \#14.5 & $(-1.37\pm0.80)\times10^{2}$\tabularnewline
\hline 
eejs15 & 64 & $3.70\times10^{-1}$ & $2.62$ & eejs15 \#14.6 & $(-1.360\pm0.038)\times10^{3}$\tabularnewline
\hline 
\hline 
eejs15 & 65 & $4.97\times10^{-1}$ & $0.538$ & eejs15 \#15.1 & $0\pm42$\tabularnewline
\hline 
eejs15 & 66 & $4.64\times10^{-1}$ & $0.748$ & eejs15 \#15.2 & $(1.40\pm0.41)\times10^{2}$\tabularnewline
\hline 
eejs15 & 67 & $6.72\times10^{-1}$ & $1.08$ & eejs15 \#15.3 & $(8.03\pm0.44)\times10^{2}$\tabularnewline
\hline 
eejs15 & 68 & $9.02\times10^{-2}$ & $1.62$ & eejs15 \#15.4 & $(4.83\pm0.63)\times10^{2}$\tabularnewline
\hline 
eejs15 & 69 & $8.73\times10^{-2}$ & $2.01$ & eejs15 \#15.5 & $(1.39\pm0.10)\times10^{3}$\tabularnewline
\hline 
\hline 
eejs15 & 70 & $4.97\times10^{-1}$ & $0.538$ & eejs15 \#16.1 & $(-8.03\pm0.43)\times10^{2}$\tabularnewline
\hline 
eejs15 & 71 & $4.64\times10^{-1}$ & $0.748$ & eejs15 \#16.2 & $(-6.63\pm0.42)\times10^{2}$\tabularnewline
\hline 
eejs15 & 72 & $6.72\times10^{-1}$ & $1.08$ & eejs15 \#16.3 & $0\pm45$\tabularnewline
\hline 
eejs15 & 73 & $9.02\times10^{-2}$ & $1.62$ & eejs15 \#16.4 & $(-3.20\pm0.64)\times10^{2}$\tabularnewline
\hline 
eejs15 & 74 & $8.73\times10^{-2}$ & $2.01$ & eejs15 \#16.5 & $(5.9\pm1.0)\times10^{2}$\tabularnewline
\hline 
\hline 
eejs15 & 75 & $1.15$ & $0.646$ & eejs15 \#17.1 & $0\pm85$\tabularnewline
\hline 
eejs15 & 76 & $6.74\times10^{-1}$ & $1.27$ & eejs15 \#17.2 & $(-1.96\pm0.91)\times10^2$\tabularnewline
\hline 
eejs15 & 77 & $7.31\times10^{-2}$ & $1.67$ & eejs15 \#17.3 & $(-5.2\pm2.7)\times10^{2}$\tabularnewline
\hline 
\hline 
ejs15 & 78 & $4.73\times10^{-1}$ & $0.506$ & ejs15 \#18.1  & $(3.04\pm0.32)\times10^2$\tabularnewline
\hline 
ejs15 & 79 & $7.93\times10^{-1}$ & $0.808$ & ejs15 \#18.2  & $(2.39\pm0.29)\times10^2$\tabularnewline
\hline 
ejs15 & 80 & $8.08\times10^{-1}$ & $1.38$ & ejs15 \#18.3 & $0\pm35$\tabularnewline
\hline 
ejs15 & 81 & $5.31\times10^{-2}$ & $2.37$ & ejs15 \#18.4 & $(-3.20\pm0.50)\times10^{2}$\tabularnewline
\hline 
ejs15 & 82 & $1.27\times10^{-1}$ & $3.04$ & ejs15 \#18.5 & $(-8.07\pm0.44)\times10^{2}$\tabularnewline
\hline 
\hline 
jsres & 83 & $1.04$ & $0.621$ & jsres \#19.1 & $0\pm14$\tabularnewline
\hline 
jsres & 84 & $1.27$ & $1.12$ & jsres \#19.2 & $113\pm17$\tabularnewline
\hline 
jsres & 85 & $1.01\times10^{-1}$ & $2.17$ & jsres \#19.3 & $(5.67\pm0.62)\times10^{2}$\tabularnewline
\hline 
jsres & 86 & $3.42\times10^{-2}$ & $2.41$ & jsres \#19.4 & $(2.47\pm0.33)\times10^{2}$\tabularnewline
\hline 
jsres & 87 & $1.48\times10^{-1}$ & $2.44$ & jsres \#19.5 & $(3.07\pm0.77)\times10^{2}$\tabularnewline
\hline 
jsres & 88 & $7.92\times10^{-2}$ & $2.57$ & jsres \#19.6 & $(7.59\pm0.52)\times10^{2}$\tabularnewline
\hline 
jsres & 89 & $1.57\times10^{-1}$ & $2.7672$ & jsres \#19.7 & $(1.74\pm0.20)\times10^{2}$\tabularnewline
\hline 
jsres & 90 & $1.70\times10^{-1}$ & $2.7654$ & jsres \#19.8 & $(5.92\pm0.62)\times10^{2}$\tabularnewline
\hline 
\hline 
jsres & 91 & $1.04$ & $0.621$ & jsres \#20.1 & $-113\pm17$\tabularnewline
\hline 
jsres & 92 & $1.27$ & $1.12$ & jsres \#20.2 & $0\pm20$\tabularnewline
\hline 
jsres & 93 & $1.01\times10^{-1}$ & $2.17$ & jsres \#20.3 & $(4.52\pm0.63)\times10^{2}$\tabularnewline
\hline 
jsres & 94 & $3.42\times10^{-2}$ & $2.41$ & jsres \#20.4 & $(1.34\pm0.34)\times10^{2}$\tabularnewline
\hline 
jsres & 95 & $1.48\times10^{-1}$ & $2.44$ & jsres \#20.5 & $(1.94\pm0.78)\times10^{2}$\tabularnewline
\hline 
jsres & 96 & $7.92\times10^{-2}$ & $2.57$ & jsres \#20.6 & $(6.45\pm0.52)\times10^{2}$\tabularnewline
\hline 
jsres & 97 & $1.57\times10^{-1}$ & $2.7672$ & jsres \#20.7 & $61\pm21$\tabularnewline
\hline 
jsres & 98 & $1.70\times10^{-1}$ & $2.7654$ & jsres \#20.8 & $(4.78\pm0.62)\times10^{2}$\tabularnewline
\hline 
	\caption*{Supplementary Information Table: {\bf The} $\mathbf \Delta^{17}{\rm O}$ {\bf differences between the 
	planets in each analyzed system and the impacted-planets.} See the main text and Methods for further details.}
\end{longtable}

\end{document}